\documentclass[twocolumn,multicolumn,aps,prb,showpacs]{revtex4}
\usepackage{graphicx}
\usepackage{dcolumn}
\usepackage{bm}
\usepackage{color}
\usepackage{latexsym}
\usepackage[T1]{fontenc}
\usepackage{amsmath}

\input{epsf}
\begin{document}

\preprint{APS/123-QED}

\preprint{APS/123-QED}

\date{\today}

\title{Disorder, critical currents, and vortex pinning energies in isovalently substituted BaFe$_{2}$(As$_{1-x}$P$_{x}$)$_{2}$ }
\author{S. Demirdi\c{s}}
\affiliation{Laboratoire des Solides Irradi\'{e}s, CNRS UMR 7642 \& CEA-DSM-IRAMIS, Ecole Polytechnique, F-91128 Palaiseau cedex, France }
\author{Y. Fasano }
\affiliation{Laboratorio de Bajas Temperaturas, Centro At\'{o}mico Bariloche \& Instituto Balseiro, Avenida Bustillo 9500, 8400 Bariloche, Argentina}
\author{S. Kasahara and T. Terashima}
\affiliation{Research Center for Low Temperature \& Materials Sciences, Kyoto University, Sakyo-ku, Kyoto 606-8501, Japan }
\author{T. Shibauchi and Y. Matsuda}
\affiliation{Department of Physics, Kyoto University, Sakyo-ku, Kyoto 
606-8502, Japan }
\author{Marcin Konczykowski}
\affiliation{Laboratoire des Solides Irradi\'{e}s, CNRS UMR 7642 \& CEA-DSM-IRAMIS, Ecole Polytechnique,  F--91128 Palaiseau cedex, France}
\author{H. Pastoriza}
\affiliation{Laboratorio de Bajas Temperaturas, Centro At\'{o}mico Bariloche \& Instituto Balseiro, Avenida Bustillo 9500, 8400 Bariloche, Rio Negro, Argentina}
\author{C.J. van der Beek}
\affiliation{Laboratoire des Solides Irradi\'{e}s, CNRS UMR 7642 \& CEA-DSM-IRAMIS, Ecole Polytechnique, F 91128 Palaiseau cedex, France }

\begin{abstract}
We present a comprehensive overview of vortex pinning in single crystals of the isovalently substituted iron-based superconductor BaFe$_{2}$(As$_{1-x}$P$_{x}$)$_{2}$, a material that qualifies as an archetypical clean superconductor, containing only sparse strong point--like pins [in the sense of C.J. van der Beek {\em et al.}, Phys. Rev. B {\bf 66}, 024523 (2002)]. Widely varying critical current values for nominally similar compositions show  that flux pinning is of extrinsic origin. Vortex configurations,  imaged using the Bitter decoration method, show less density fluctuations than those previously observed in charge-doped  Ba(Fe$_{1-x}$Co$_{x}$)$_{2}$As$_{2}$ single crystals. Analysis reveals that the pinning force and -energy distributions depend on  the P-content $x$. However, they are always much narrower than in Ba(Fe$_{1-x}$Co$_{x}$)$_{2}$As$_{2}$, a result that is attributed to the weaker temperature dependence of the superfluid density on approaching $T_{c}$ in BaFe$_{2}$(As$_{1-x}$P$_{x}$)$_{2}$. Critical current density measurements and pinning force distributions independently yield a mean distance between effective pinning centers $\overline{\mathcal L} \sim 90$~nm, increasing with increasing P-content $x$. This evolution can be understood as being the consequence of the P-dependence of the London penetration depth.  Further salient features are a wide vortex free ``Meissner belt'', observed at the edge of overdoped crystals, and characteristic chain-like vortex arrangements, observed at all levels of P-substitution.
\end{abstract}

\maketitle

\section{Introduction}

Recent vortex imaging studies  performed on iron pnictide superconductors  show  evidence for nanoscale inhomogeneity \cite{demirdis,Kees}  being at the origin of the low-field critical current density and the highly disordered vortex structures in these materials. \cite{Eskildsen,Eskildsen2009,Inosov,Vinnikov,Kalisky,Luan,Yi Yin,haihuwen,hazuki,furukawa} Notably, in Ba(Fe$_{1-x}$Co$_{x}$)$_{2}$As$_{2}$, the critical current density $j_c$ and vortex distributions imaged by Bitter decoration could be consistently analyzed, provided that  spatial heterogeneity, on a scale of several dozen nm,  both of the critical temperature $T_c$ and the vortex line energy $\varepsilon_0$,  is taken to be responsible for flux pinning.\cite{demirdis}  At higher magnetic  fields, of the order of several tenths of Tesla, nano-scale heterogeneities are inefficient in pinning flux lines.\cite{Kees,Kees1,Kees2} The critical current density is then most likely determined\cite{Kees2}  by the scattering of quasiparticles in the vortex cores associated with the presence of atomic-size defects in the crystal,\cite{Thuneberg84,Blatter94} leading to weak collective pinning.\cite{Blatter94,Larkin79} A good candidate for these defects are the dopant atoms themselves.\cite{demirdis,Kees,Kees1,Kees2} The nature of the dopant atoms is essential for this mechanism; charged defects lead to different scattering than  uncharged defects. \cite{Kees2} This weak  collective pinning contribution to the critical current density manifests itself as a plateau-like behavior in a $j_{c}(B)$ plot. It is  present in all charge--doped iron-based superconductors, as well as in Ba(Fe$_{1-x}$Ru$_{x}$)$_{2}$As$_{2}$.\cite{creep} 

On the other hand, in isovalently substituted BaFe$_{2}$(As$_{1-x}$P$_{x}$)$_{2}$  there is no indication of weak collective pinning,\cite{Kees2} which qualifies the material as ``clean'' with respect to charge-doped iron-based superconductors. Given recent claims \cite{haihuwen,furukawa} that in certain iron-based superconductors, the vortex configuration is more ordered than what was hitherto observed,\cite{Eskildsen,Eskildsen2009,Inosov,Vinnikov,Kalisky,Luan,Yi Yin,hazuki} it is interesting to see whether the absence of  weak collective pinning has any impact on the spatial configuration of vortices.  From a magnetic force microscopy (MFM) study at magnetic fields up to 100 Oe, Yang {\em et al.} claim that vortex configurations in hole-doped single crystalline Ba$_{1-x}$K$_x$Fe$_2$As$_2$ (with $x=0.28$ and $x=0.4$) are more ordered  than in, {\em e.g.} Ba(Fe$_{1-x}$Co$_{x}$)$_{2}$As$_{2}$. \cite{demirdis} 
Using an analysis method similar to that of Ref.~\onlinecite{demirdis},  they report pinning forces that are one order of magnitude smaller on average. They also claimed the observation of local triangular vortex order in the optimally doped material, even though closer scrutiny (see section~\ref{section:Bitter} below) reveals the vortex ensembles to be no more ordered than those in Ba(Fe$_{0.925}$Co$_{0.075}$)$_{2}$As$_{2}$.\cite{demirdis}  Finally, the authors \cite{haihuwen} reported on the observation of remarkable vortex chains, both in underdoped ($x = 0.28$) and optimally doped ($x=0.4$) Ba$_{1-x}$K$_x$Fe$_2$As$_2$. The presence of these was attributed to vortex pinning by the twin boundaries arising from the orthorhombic structure, at least in the underdoped material.  
Furthermore, neutron scattering experiments on the vortex lattice in isovalently substituted BaFe$_{2}$(As$_{1-x}$P$_{x}$)$_{2}$ were performed at $T = 2$~K by Kawano-Furukawa {\em et al.}.\cite{furukawa} No vortex Bragg peaks were found for the optimally substituted compound. However, after annealing the samples at $500^{\circ}\mathrm{C}$, a distorted triangular vortex lattice was observed; this became more ordered as  the applied magnetic field was increased from 0.7  to 7 T. \cite{furukawa} These results suggest that the disorder responsible for pinning in BaFe$_{2}$(As$_{1-x}$P$_{x}$)$_{2}$ is extrinsic in nature. 

In this work we present and analyze sustainable current density measurements, magneto-optical imaging, and Bitter decoration experiments performed on BaFe$_{2}$(As$_{1-x}$P$_{x}$)$_{2}$ single crystals with different $x$. In contrast to some reports,\cite{Fang} we find strong pinning, presumably by nanoscale heterogeneity, as the only observed pinning contribution in fields up to 5~T. As in Ba(Fe$_{1-x}$Co$_{x}$)$_{2}$As$_{2}$ and other materials,\cite{Vinnikov} there is no evidence for any extended triangular order in the vortex ensemble; thus, the strong pinning contribution in itself suffices to generate the extreme disorder of the vortex ensemble. The spatial configuration of vortices in isovalently substituted BaFe$_{2}$(As$_{1-x}$P$_{x}$)$_{2}$ does not present large vortex density fluctuations such as observed in charge-doped Ba(Fe$_{1-x}$Co$_{x}$)$_{2}$As$_{2}$ single crystals,\cite{demirdis} a fact that is attributed to the different temperature dependences of the superfluid density in the two materials. The quantitative analysis of the vortex configurations in terms of the pinning energy confirms that pinning disorder is somewhat less effective in  BaFe$_{2}$(As$_{1-x}$P$_{x}$)$_{2}$ than in Ba(Fe$_{1-x}$Co$_{x}$)$_{2}$As$_{2}$, and that it depends on the P-content $x$. Analysis of the pinning energies, pinning forces, and the critical current density as function of P-content yields consistent estimates of the effective pin density. This clearly decreases upon increasing the P-content, a behavior that tracks  the composition dependence of the scattering rate in the normal state. The main features of our results can be understood in terms of the composition dependence of the vortex line energy, implying that local variations of the superfluid density is a good candidate for the origin of the vortex pinning. The local variation of the dopant atom density as well as the variation of the sustainable current density with composition argues against any possible spatially phase--separated superconducting and anti-ferromagnetic states of the material as being at the origin of pinning.

\section{Experimental details}

Experiments have been performed on BaFe$_{2}$(As$_{1-x}$P$_{x}$)$_{2}$ single crystals grown by the self-flux method,\cite{Pdoped} and  characterized using Energy Dispersive X-Ray Spectroscopy (EDX) and EDX mapping in a scanning electron microscope (SEM).  Crystals with manifest chemical heterogeneity were discarded from further study.  The crystals described below present no impurity phases, within the experimental limits of accuracy $\lesssim$ 1$\%$. 

Magnetic flux penetration in crystals with different substitution levels  ($x = 0.27-0.49$) was characterized using the magneto-optical imaging (MOI) method\cite{Kees,Dorosinskii,Kees1}  before further experiments. The MOI technique notably allows one to discard samples with macroscopic defects, and also, to extract calibrated flux density profiles. The sustainable current density $j_{s}$  for magnetic fields up to $\mu_{0}H_{a}=500$\,G (50~mT) was  obtained from  the gradient of the local  magnetic  flux  density $B$ perpendicular to the crystal surface,  using the Bean model.\cite{Zeldov,Bean} Given the thickness-to-width ratio of these crystals, $d/w \sim 0.25$, one has $\mu_{0} j_{s} \sim 3 \,  dB/dx$.\cite{brandt} The crystal inhomogeneity, and notably  the local distribution of $T_c$ was characterized using the differential magneto-optical (DMO) method. \cite{Soibel,Kees1} Measurements in higher magnetic fields were performed using micron-sized Hall probe arrays,\cite{marcin}  tailored in a pseudomorphic GaAlAs/GaAs heterostructure, as well as using a Superconducting Quantum Interference Device (SQUID)-based magnetometer. 

The vortex ensembles in several crystals, of substitution levels  $ x = 0.33$, 0.36, and 0.49, were imaged using the Bitter decoration method,\cite{demirdis,fasano} at an applied field $\mu_{0}H_{a} = 20$~G (2~mT) (see Section~\ref{section:Bitter}). In what follows, individual crystals will be identified as ( $x = $ substitution level, sample number $\#$ ).

\begin{figure}[t]
\includegraphics[width=0.4\textwidth]{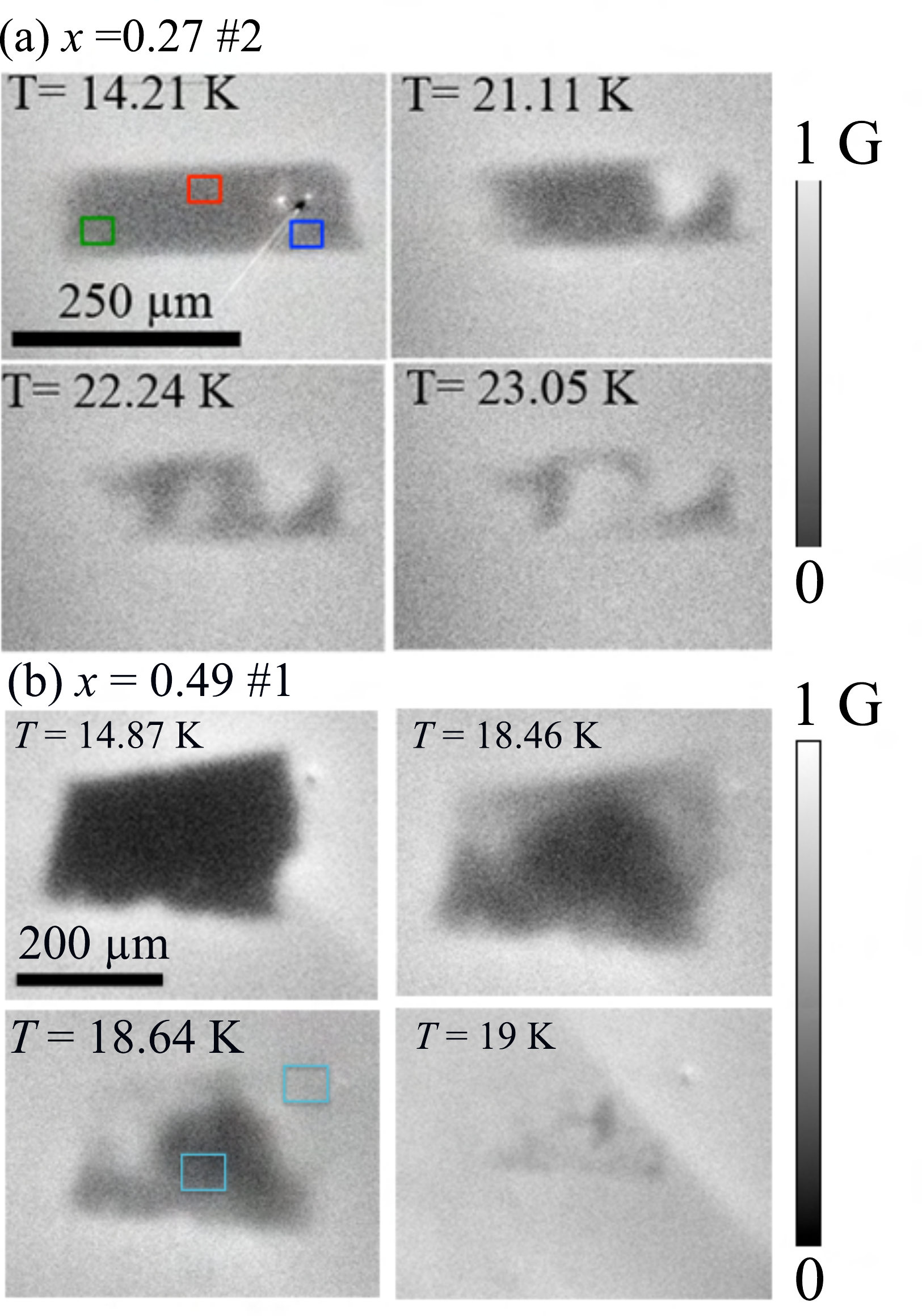}
\caption{(Color online)  Differential magneto-optical images of the screening of a magnetic field  $\mu_{0}H_{a}=1$ G (0.1~mT) by BaFe$_{2}$(As$_{1-x}$P$_{x}$)$_{2}$ single crystals  (a) $x= 0.27$\# 2, and (b) $x=0.49$\#1. The intensity  is proportional to the local magnetic flux density $B({\mathbf r})$. Thus, black areas are regions of excluded flux, while the light areas surrounding the crystal are traversed by the applied magnetic field. Rectangular frames in (a) indicate the regions where the transmittivity data of Fig.~\ref{Tc}\,(b) are determined.  The frames in (b)  denote the areas where the decoration images of Fig.~\ref{deco} (b,d) were obtained. }
\label{DMO}
\end{figure}

\section{Results}
\subsection{Spatial variation of the critical temperature $T_{c}$}
\label{currentdensity}

Figures~\ref{DMO} (a,b) presents DMO images of the exclusion of an applied field $\mu_{0}H_{a}= 1$ G (0.1~mT)   as one crosses the superconducting to normal transition of  BaFe$_{2}$(As$_{1-x}$P$_{x}$)$_{2}$ single crystals  $x=0.27 \, \#2$ and  $x=0.49 \, \#1$. These images reveal that  $T_c$ is spatially heterogeneous. While inhomogeneity is especially pronounced in the underdoped samples, see crystal  $x=0.27 \,\#$2, it is also observed in the overdoped crystals.  A link between the heterogeneity observed in underdoped samples and a possible phase separation between of superconducting- and  Spin-Density Wave (SDW) antiferromagnetic phases,  and/or  that of orthorhombic and tetragonal structural domains is therefore not obvious. 

\begin{figure}[t]
\includegraphics[width=0.6\textwidth]{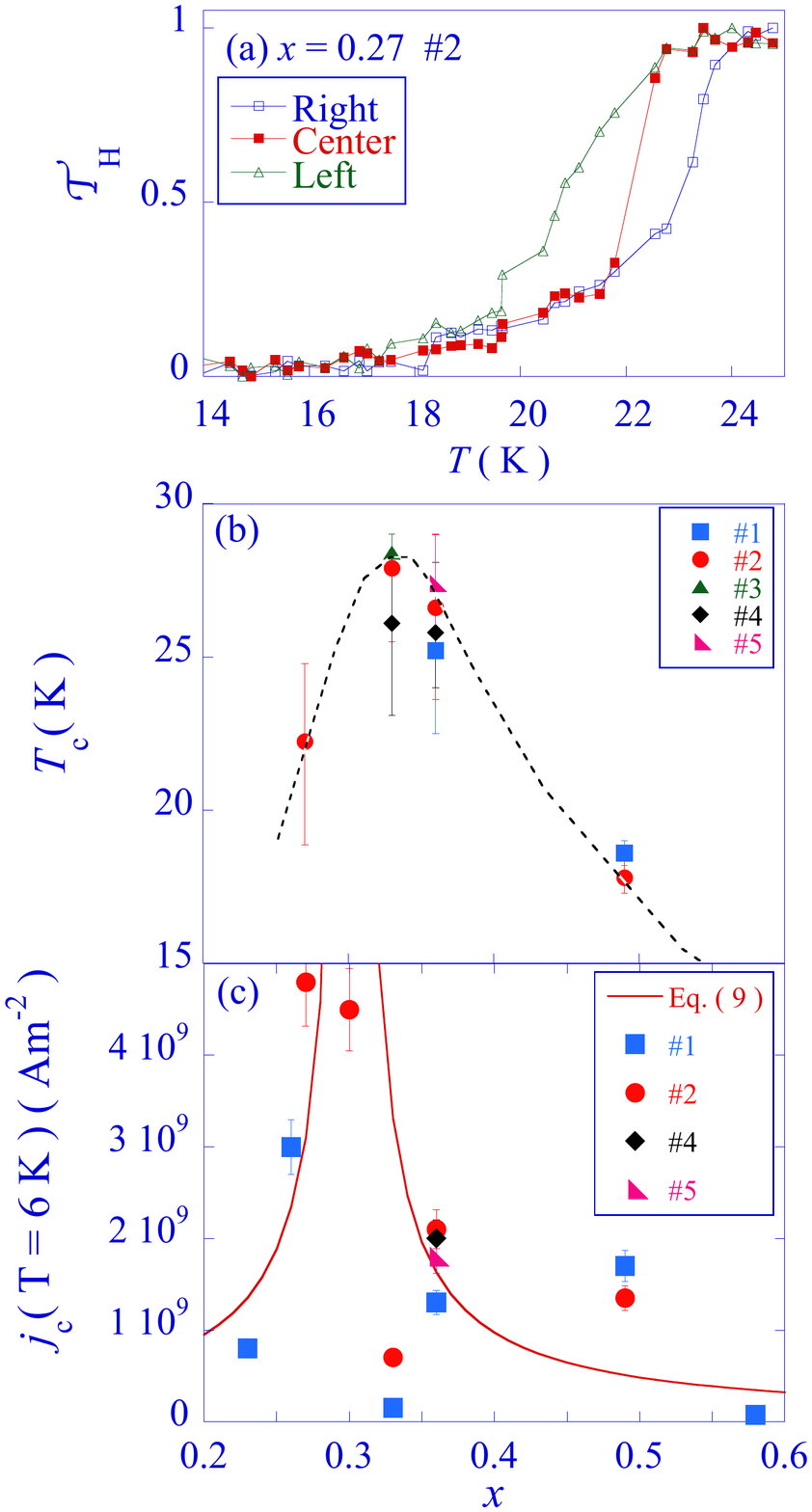}\vspace{-15mm}
\caption{(Color online) (a) $\mathcal{T}_{H}$ measured on the three regions of  crystal $x=0.27\, \#2$ indicated in Fig.~\ref{DMO}. (b) Transition temperature $T_{c}$  versus P--content $x$. The error bars indicate the local spread of $T_c$ inside a given crystal. For each $x$, the numbering  \#1, \#2, etc  denotes different crystals from the same batch. (c) Dependence of the low-temperature ($T = 6$~K) critical current density on P-content $x$. The drawn line shows the  evolution of the critical current due to spatial variations of the dopant atom density (on a scale of $\delta z \sim 100$~nm and with variance $\Delta x \sim 0.3$~\%) such as expected from Eq.~(\protect\ref{eq:dopant-density-variations}).} 
\label{Tc}
\end{figure}

The $T_{c}$--heterogeneity is quantified by the local transmittivity  $\mathcal{T}_{H} = \left[ I( \mathbf r,T ) - I( \mathbf r,T \ll T_{c} ) \right]  /$ $  \left[ I( \mathbf r,T \gg T_{c} )-I( \mathbf r,T \ll T_{c}) \right]$, extracted from the luminous  intensity $I(\mathbf r,T)$  in the DMO images of Fig.~\ref{DMO}.  $\mathcal{T}_{H}(T)$ is presented in Fig.~\ref{Tc}a for crystal $x=0.27 \, \# 2$.  The width of the superconducting transition obtained for the various crystals is presented in Fig.~\ref{Tc}\, b, where the error bar indicates the spread of $T_c$  in a given crystal, and the data points give the temperature where 50\% of the crystal has become superconducting. \vspace{-5mm}

 \begin{figure}[t] 
\includegraphics[width=0.45\textwidth]{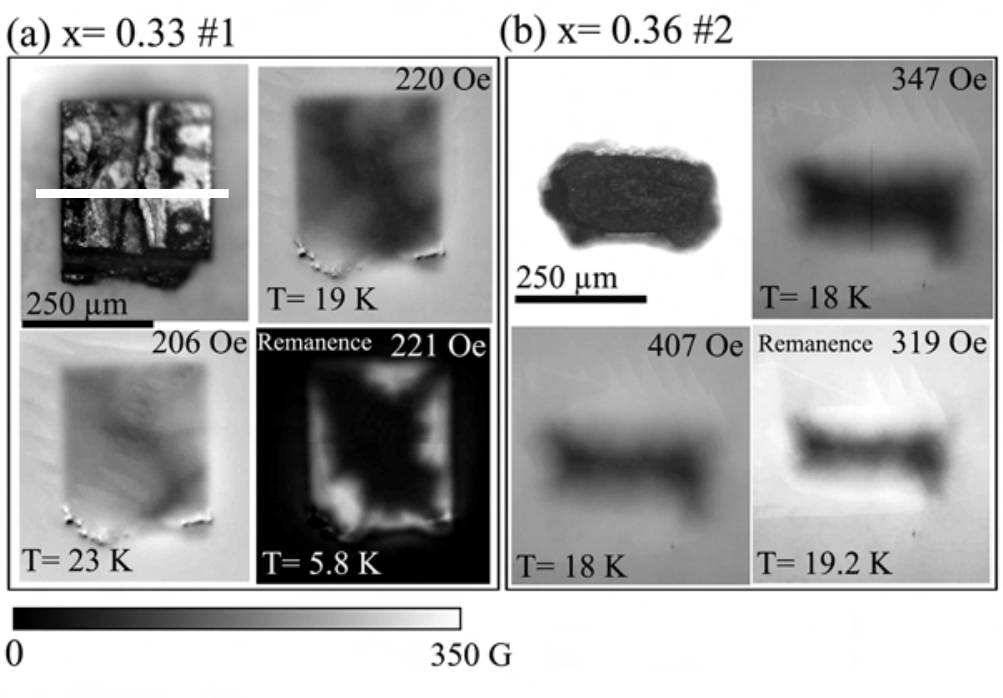}
\caption{Magnetic flux density distribution in BaFe$_{2}$(As$_{1-x}$P$_{x}$)$_{2}$ single crystals $x=0.33 \, \#$1  (a), and $ x=0.36 \,\#$2 (b), after zero--field cooling to the indicated temperatures and the application of different magnetic fields. The top left panel of each subfigure shows the respective crystal, the white lines are those along which the profiles in Fig.~\ref{profile} are extracted. The intensity in the other panels reflects the local flux density $B({\mathbf r})$. The bottom right-hand panel of (a) shows the trapped flux distribution in crystal  $x=0.33 \, \#$1, after application and removal of $\mu_{0}H_{a} = 22.1$~mT. }
\label{MOI}
\end{figure}

 \begin{figure}[b] \vspace{-4mm}
 \includegraphics[width=0.55\textwidth]{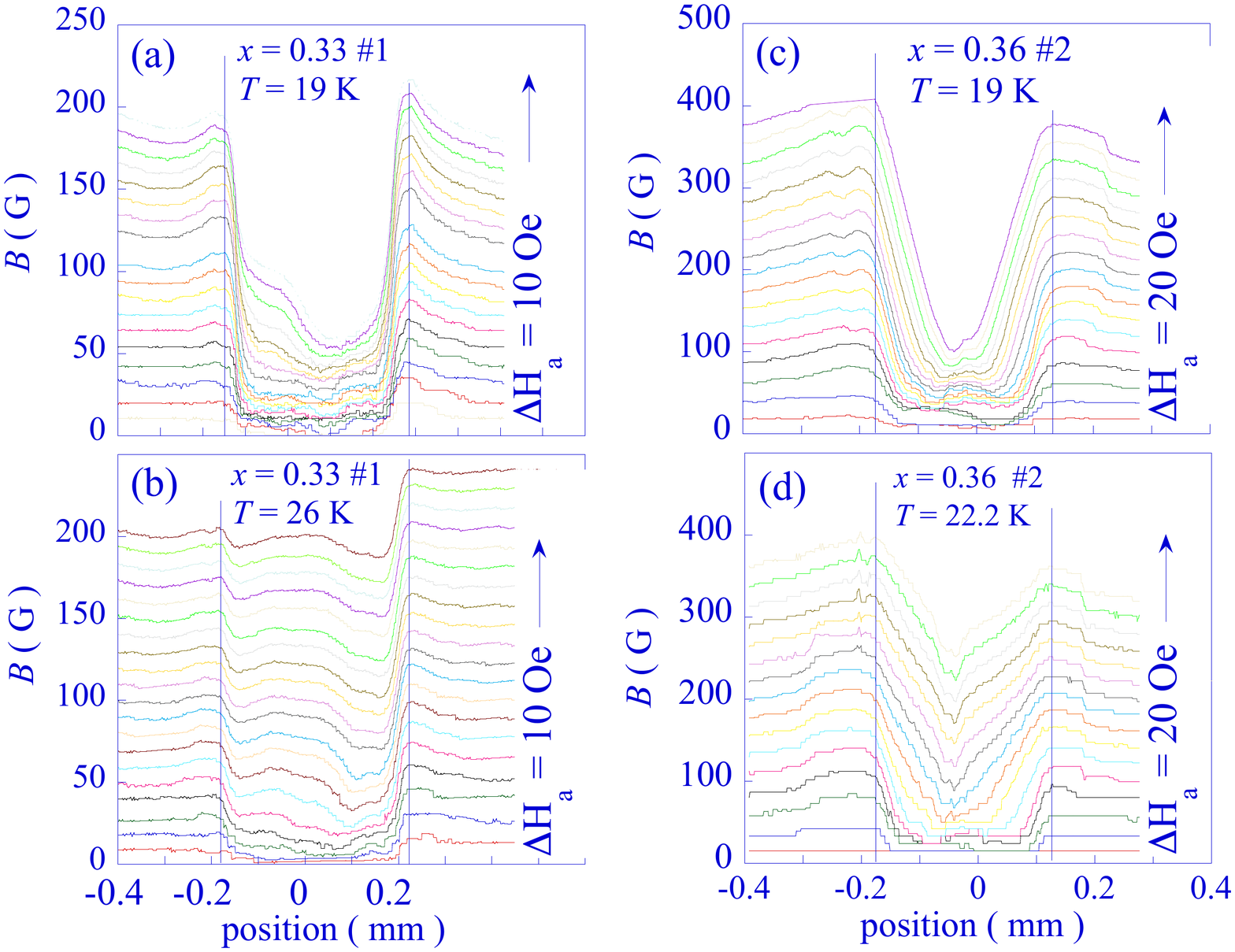}
\caption{(Color online) Magnetic flux  density profiles in BaFe$_{2}$(As$_{1-x}$P$_{x}$)$_{2}$  crystal $x=0.33 \, \# 1$, at  $T= 19$ and 26~K  (a,b), and crystal $x = 0.36 \, \# 2$, at $T=19$ and 22.2~K (c,d), after zero-field cooling and application of the magnetic field in successive steps $\Delta H_{a}$.  The Bean--like profiles in (c,d) are obtained from the MOI images of Fig.~\protect\ref{MOI}(b). The  profiles  in (a,b) are influenced by a surface barrier and correspond to the crystal of Fig.~\protect\ref{MOI}(a). }
\label{profile}
\end{figure}

\begin{figure}[h]
 \includegraphics[width=0.63\textwidth]{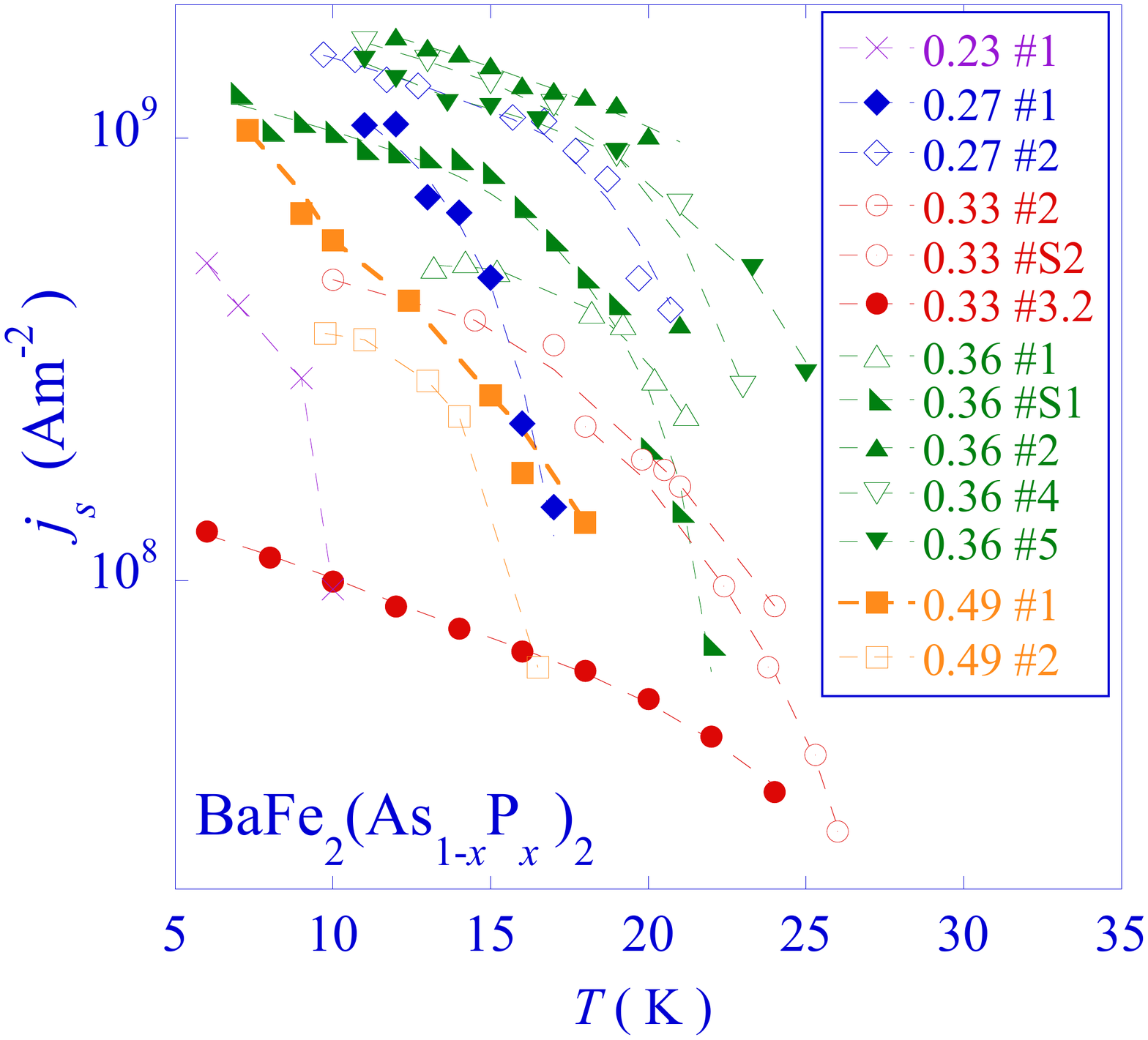}
\caption{(Color online) Temperature dependence of the sustainable current density $j_{s}(T, B=30\,\mathrm{mT})$, determined from  flux density profiles obtained from MOI images, for a variety of crystals with substitution levels $0.23 \leq x \leq 0.49$. Samples denoted ``S'' concern a limited area of the sample with the same number. Dashed lines are guides to the eye.}
\label{Jc}
\end{figure}
 
\begin{figure}[ht]
\includegraphics[width=0.63\textwidth]{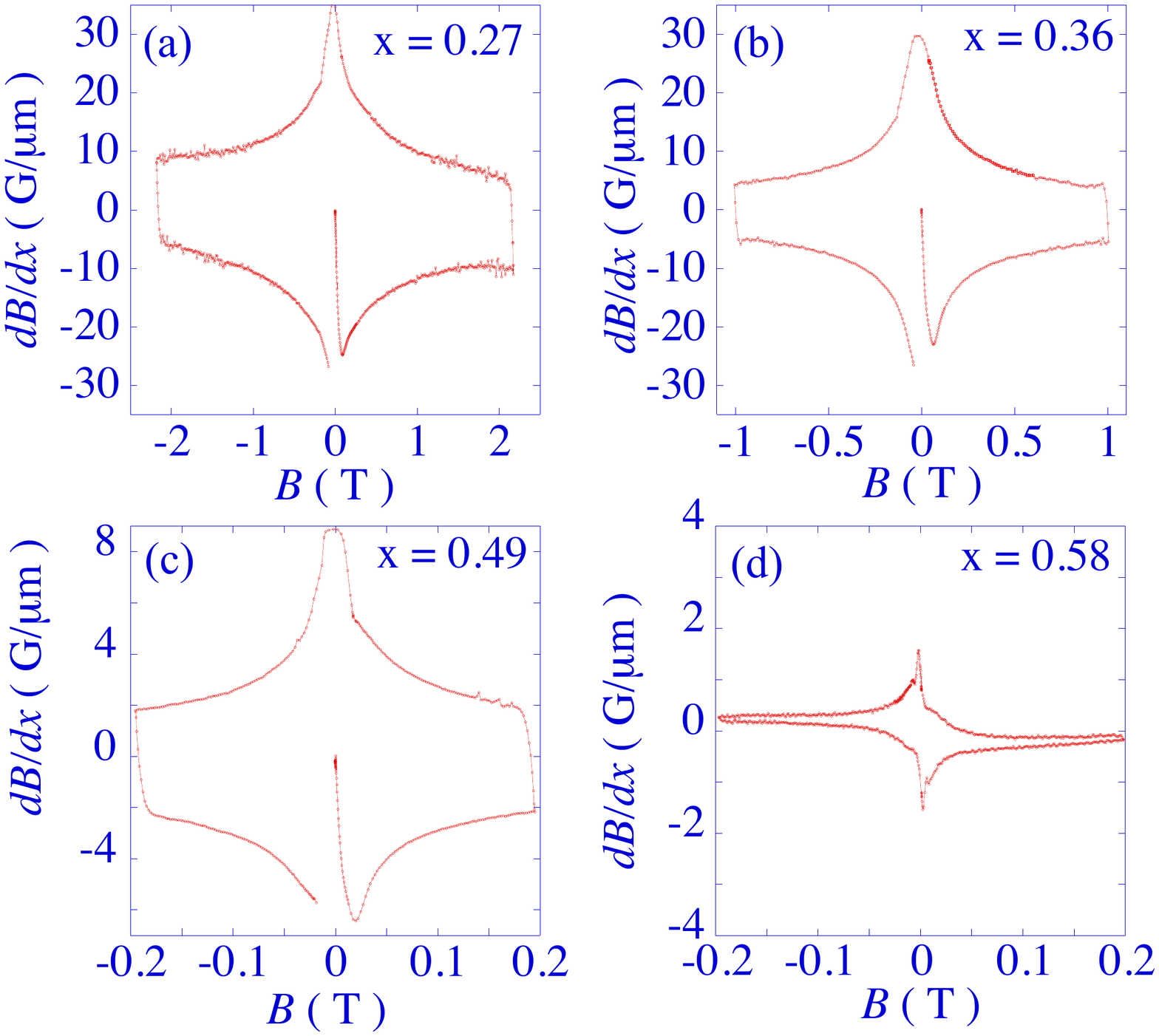}
\caption{(Color online) Hysteresis loops of the spatial gradient $dB_{}/dx$ of the local magnetic induction on the surface of BaFe$_{2}$(As$_{1-x}$P$_{x}$)$_{2}$ single crystals of different substitution levels $0.27\leq x \leq 0.58$, measured using the Hall--probe magnetometry technique, \protect\cite{marcin} at $T= 6$~K.}
\label{MH}
\end{figure}

 \begin{figure}[b]
  \vspace{-4mm}
\includegraphics[width=0.66\textwidth]{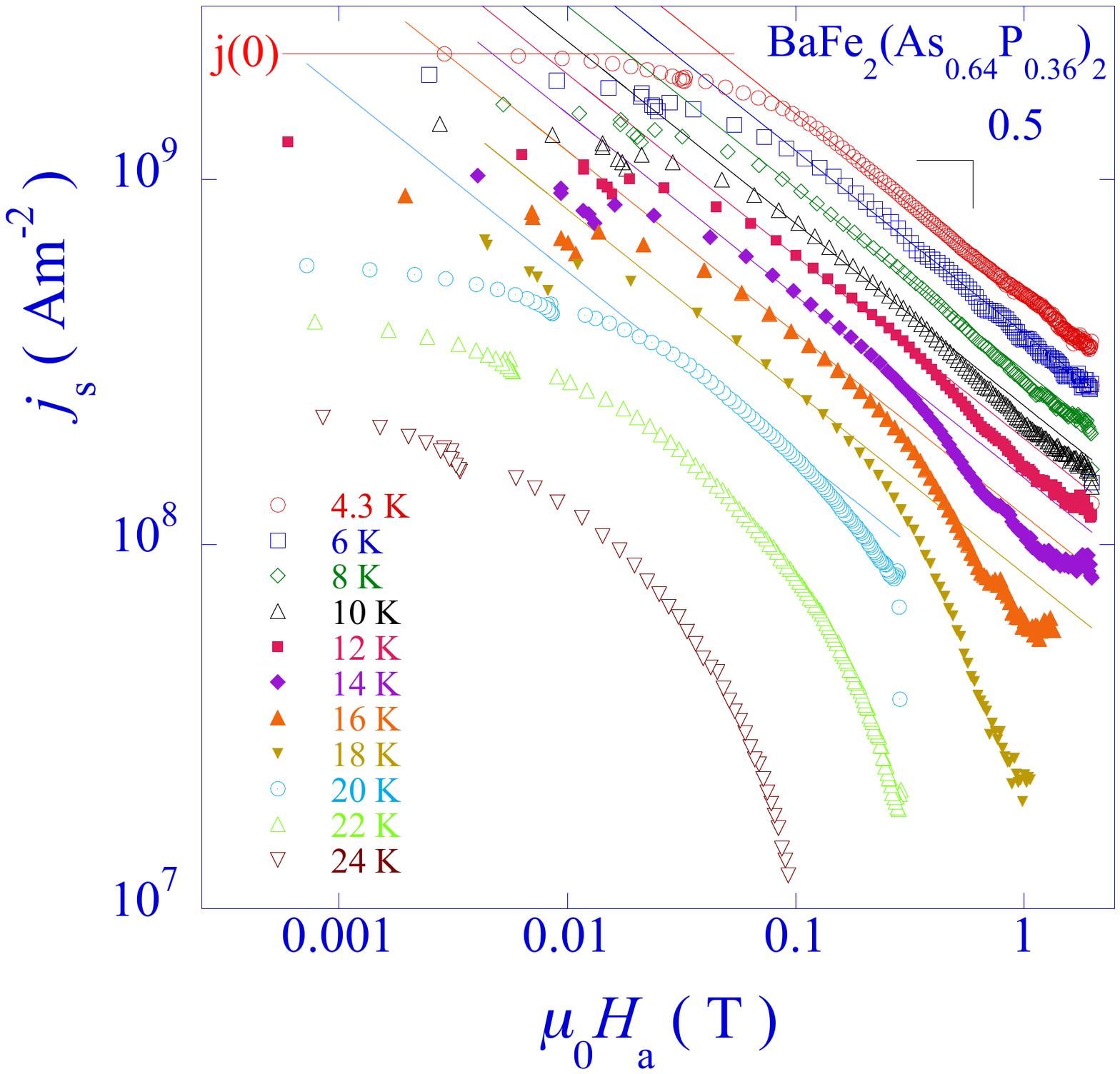}
  \vspace{-10mm}
\caption{(Color online) Magnetic field dependence of  the sustainable current density $j_{s}$ for crystal $x=0.36$ \# 2, obtained from the local magnetic induction gradient as measured using the Hall--probe magnetometry method.\protect\cite{marcin}}
\label{j(b)}
\end{figure}
  
  \begin{figure}[ht]
\includegraphics[width=0.67\textwidth]{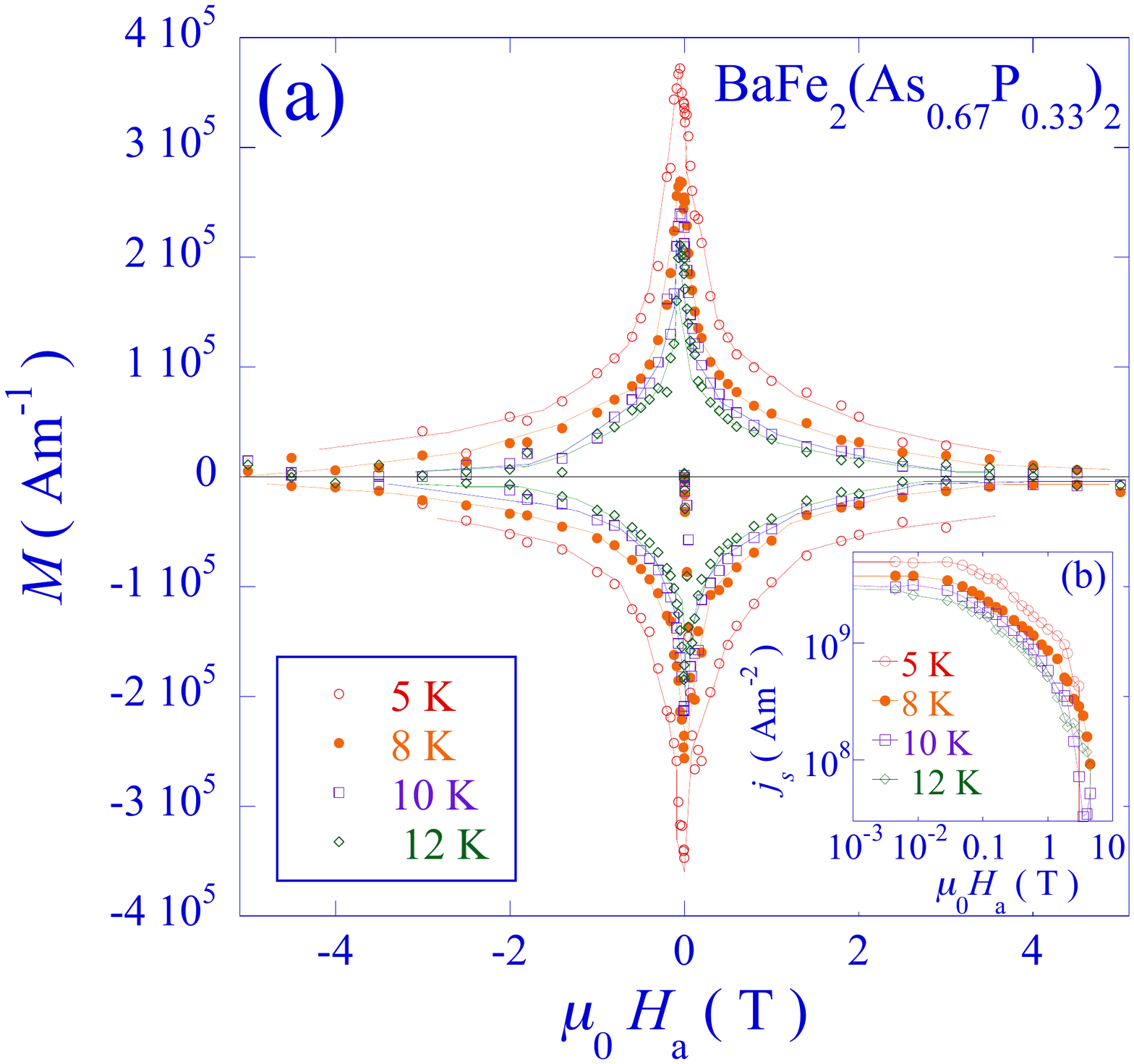}
\vspace{-6mm}
\caption{(Color online) (a) Magnetic hysteresis loops measured on  BaFe$_{2}$(As$_{1-x}$P$_{x}$)$_{2}$  crystal $x=0.33 \, \# 2$ using a SQUID magnetometer. (b) Sustainable screening current $j_{s}$ as function of the applied field $\mu_{0}H_{a}$(in Tesla). }
\label{SQUID}
\end{figure}

 \subsection{Sustainable current density $j_{s}$}
 
Figure~\ref{MOI} presents MOI of the magnetic flux penetration (after zero-field cooling) into superconducting BaFe$_{2}$(As$_{1-x}$P$_{x}$)$_{2}$ single crystals 
$x=0.33 \, \# 1$, and $x=0.36 \, \# 2$, respectively. The former crystal is characterized by very weak bulk pinning and, as a result, a large influence of  geometrical \cite{Zeldov94,Brandt99} and surface barriers.\cite{Livingston64} The influence of the surface screening current leads to an inhomogeneous flux density distribution, as  presented in Figure~\ref{MOI}\,(a). In contrast, crystal $x=0.36 \, \#2$ shows regular flux penetration, in accordance with the Bean critical state model. The influence of a surface barrier, present for both flux entry and flux exit, is also revealed by  Hall--probe array measurements.

The flux density profiles across the same crystals,
depicted in Figure~\ref{profile} (a,b) and (c,d) respectively, were extracted from the calibrated luminous intensities of the magneto-optical images in Fig.~\ref{MOI}. One sees that, even for the same or comparable doping levels, very different flux density profiles can be obtained after zero-field cooling. Figure ~\ref{profile}\, (c,d) shows the Bean-like penetration of the magnetic flux inside crystal $x=0.36 \, \# 2$, with no clear influence of a surface barrier, while the flux profiles for crystal  $x=0.33 \, \# 1$ in Figure~\ref{profile}\,(a,b) show, apart from inhomogeneity, a large discontinuity in the magnetic induction at the sample edge, characteristic of a surface barrier.  Given the very different behavior for nearly the same sample composition,  the origin of the bulk critical current density in BaFe$_{2}$(As$_{1-x}$P$_{x}$)$_{2}$ is most likely extrinsic. This is supported by the temperature dependence of the sustainable current density $j_{s}(T,B=30\, \mathrm{mT})$ of the studied samples, shown in Fig.~\ref{Jc}. The absolute value of $j_{s}(T)$ is widely disperse, even for crystals with the same doping level. 

In spite of the disparity, the flux pinning mechanism in all crystals is the same. Fig.~\ref{MH} shows hysteresis loops of the local gradient of the magnetic induction $dB_{}/dx$ in fields of up to 2~T, obtained on crystals of different composition using the Hall probe--array magnetometry technique.\cite{marcin} The hysteresis loops  were measured at 6 K, at which flux creep has only a moderate influence. For all crystals, of all investigated substitution levels, one has the ubiquitous central peak at zero field, believed to be due to strong pinning by nm-scale disorder.\cite{Kees1} The magnetic field dependence of the sustainable current density $j_{s}(B)$ was obtained from the value of $dB_{}/dx$ at given $B$. Fig.~\ref{j(b)} shows  $j_{s}(B)$ for the optimally substituted single crystal $x=0.36$ \# 2 at different temperatures. The $j_{s}(B)$ curve at the lowest $T$  is representative of the field-dependent critical current density $j_c(B)$. The critical current density is characterized by a low--field plateau, followed by a $j_{c} \propto B^{-1/2}$ decrease at higher fields, behavior that is typical of strong flux pinning by sparse point-like defects.\cite{Kees1,Kees2,vdBeek2002} The contribution to $j_{c}(B)$ due to weak collective pinning of the vortex lines by atomic sized point pins,  observed in all charge-doped iron based superconductors as well as in Ba(Fe$_{1-x}$Ru$_{x}$)$_{2}$As$_{2}$, is clearly absent in  Fig.~\ref{j(b)}. The BaFe$_{2}$(As$_{1-x}$P$_{x}$)$_{2}$ system can therefore be seen as a typical strongly pinning superconductor, in the sense that only large but sparse extrinsic point--like pins contribute to flux pinning.\cite{vdBeek2002} At larger temperatures and fields,  $j_{s}(B)$ decreases faster than $B^{-1/2}$, an effect attributed to flux creep. 

Measurements carried out to larger fields using a SQUID magnetometer show that, contrary to the data presented in Ref.~\onlinecite{Fang}, the sustainable current  density continues its monotonous decrease as function of magnetic field -- there is no  ``fishtail'' or ``second-peak''  effect\cite{Kees1} up to $\mu_{0}H_{a} = 5$~T. The peak effect being usually associated with a weak collective pinning contribution to the critical current, we surmise that in the field range of interest the strong pinning mechanism is the only one at play.

\begin{figure}[t]
\includegraphics[width=0.5\textwidth]{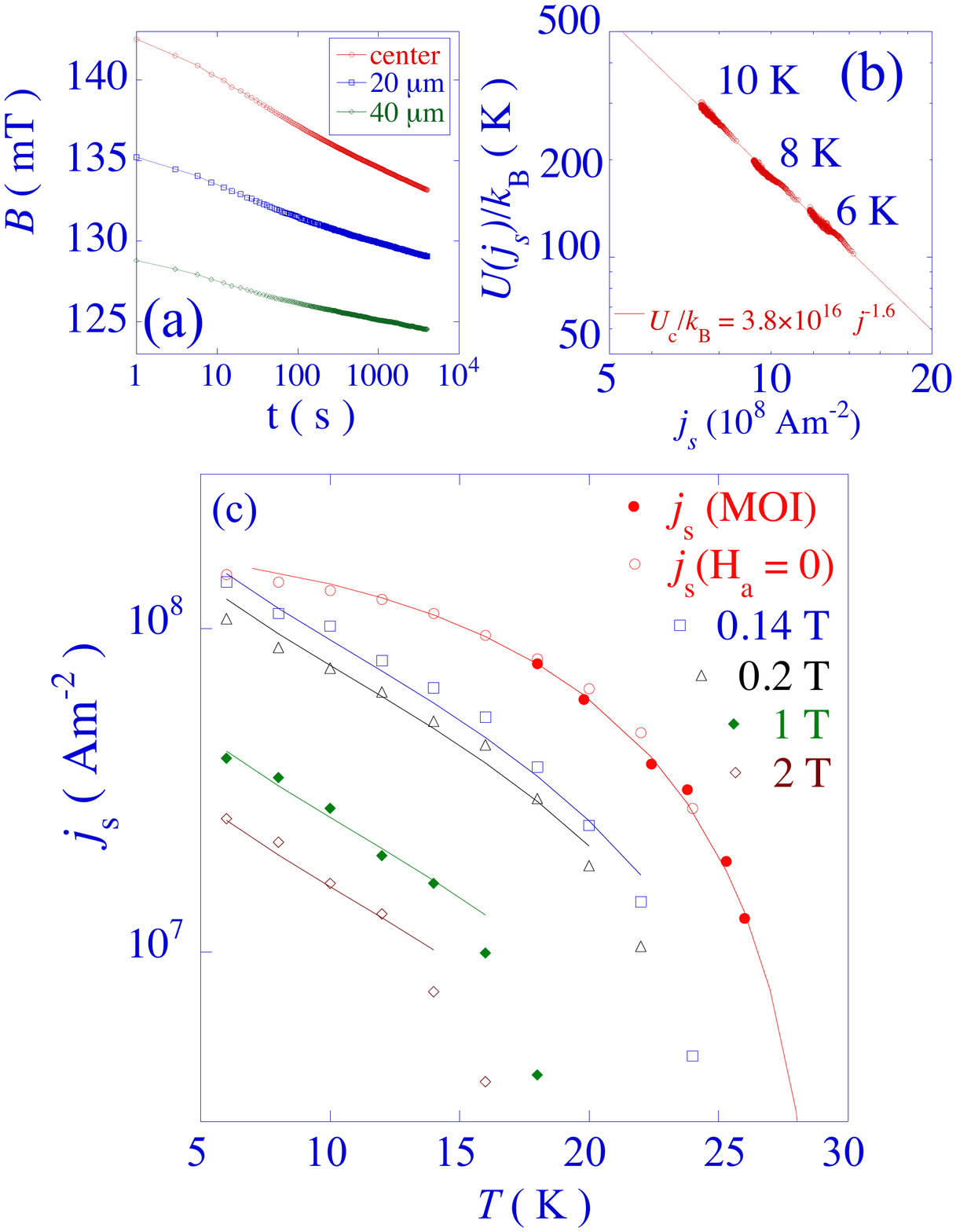}
\vspace{-20mm}
\caption{(Color online) Magnetic relaxation in BaFe$_{2}$(As$_{1-x}$P$_{x}$)$_{2}$  crystal $x=0.33 \, \# 2$. (a)  Relaxation of the magnetic flux density at three positions on the crystal surface -- the center, and positions 20 and 40~$\mu$m from the center. Data were taken at 10~K, after field cooling in 0.2~T and subsequently removing the  field. (b) Experimental flux-creep activation barrier $U$ versus $j_{s}$, as obtained using the method outlined in Refs.~\onlinecite{Abulafia95,Konczykowski2012}. (c) Temperature dependence of the sustainable current density in zero applied field, and applied fields of 0.14, 0.2, 1, and 2~T. Measurements using the Hall probe-array technique (open symbols) were obtained from hysteresis loops such as shown in Fig.~\protect\ref{MH}, while data from MOI are obtained from the flux-profile gradient. The drawn line through the $j_{s}(H_{a} = 0)$--data  corresponds to the expected $T$-- dependence of the depairing current; the drawn lines for higher fields are obtained from this by taking the field--dependence of the critical current  (\protect\ref{current1},\protect\ref{current}) into account, and by correcting for flux creep using Eq.~(\protect\ref{eq:creep}).   }
\label{fig:creep}
\end{figure}

\subsection{Quantitative effect of flux creep}

The influence of flux creep is assessed from relaxation measurements of the local flux density using the Hall probe magnetometry technique. Typical examples, shown in Fig.~\ref{fig:creep}\,(a), show that the creep rate $S \equiv d\ln (dB_{}/dx) / d \ln t$ typically amounts to a few percent. Nevertheless, $j_{s}$ is significantly affected by creep,  such that it is determined as the solution of the relation $U(j_{s})  = k_{B}T \ln \left[ \left( t_{0}  + t \right) / \tau \right]$, rather than by the critical current density $j_{c}$.\cite{Geshkenbein90,vdBeek92} Here, $t_{0}$ is a time describing transient effects at the onset of relaxation, and $\tau$ is a normalization time related to the sample inductance.\cite{Geshkenbein90,vdBeek92}  The dependence of the flux creep barrier $U(j)$ on current density $j$ can be extracted using various methods, including those of Maley {\em et al.} \cite{Maley90} and Abulafia {\em et al.}.\cite{Abulafia95,Konczykowski2012} Applying the latter, we find [see Fig.~\ref{fig:creep}(b)] that the creep barrier in optimally substituted BaFe$_{2}$(As$_{1-x}$P$_{x}$)$_{2}$ follows
\begin{equation}
U(j) = U_{c} \left( \frac{j_{c}}{j}\right)^{\mu},
\end{equation}
with values of the exponent $\mu \sim 1.5 - 2$. Therefore, the time and temperature-dependence of the sustainable screening current density is described by 
\begin{equation}
j_{s} = j_{c}\left[ \frac{k_{B}T}{U_{c}} \ln\left( \frac{t + t_{0}}{\tau} \right)\right]^{-1/\mu}.
\label{eq:creep}
\end{equation}
The impact of flux creep on the temperature dependence of the sustainable current is depicted in Fig.~\ref{fig:creep}(c), which shows $j_{s}(T)$--curves for crystal $x = 0.33$ \#2, for different $B$. The curve in zero applied field is little affected by creep, and roughly follows the expected temperature dependence of the depairing current, $j(0,T) \sim \varepsilon_{0}(T)/ \Phi_{0}\xi(T)$  ($\Phi_0 = h/2e$ is the flux quantum). Here, the vortex line energy,  $\varepsilon_{0}(T) =\Phi_{0}^{2}/4\pi\mu_{0}\lambda_{ab}^{2}$, proportional to the superfluid density $n_{s} \sim \lambda_{ab}^{-2}$, is evaluated using the data for the in--plane penetration depth $\lambda_{ab}(T)$ of Ref.~\onlinecite{Hashimoto}, and the coherence length  $\xi(T) \sim \xi(0)\sqrt{(1+T/T_{c})/(1-T/T_{c})}$. The curves for varying applied field can then be well described by taking the creep barrier prefactor $U_{c}(T)\propto \varepsilon_{0}(T)$, $j_{c}(B_{},T) \propto j(0,T) B_{}^{-1/2}$, $\mu = 1.6$, and $ \ln\left[\left( t + t_{0}\right) / \tau \right] = 20$.\cite{Konczykowski2012} Therefore, the temperature dependence of the screening current in fields larger than 0.1 T is essentially determined by flux creep.

\subsection{Extraction of pinning parameters }
We now analyze the $j_{s}(B)$--curves measured at low temperature, representative of the critical current density $j_{c}(B)$, and which bear the hallmarks of strong pinning.  These are the plateau at low magnetic field,\cite{Kees1,vdBeek2002}
\begin{equation}
j_{c}(0)  = \frac{f_{p}}{\Phi_{0}\overline{\mathcal{L}}} =  \pi^{1/2} \frac{f_{p}}{\Phi_{0}\varepsilon_{\lambda}}
\left( \frac{U_{p} n_{i}}{{\varepsilon}_{0}}\right)^{1/2} , \hspace{2mm}(B \ll B^{*})
\label{current1}
\end{equation}
followed by a power-law decrease  as a function of 
the flux density $B$,\cite{vdBeek2002,Kees1} which can be described  as
\begin{eqnarray}
j_{c}(B)   & = &  \frac{f_{p}}{\Phi_{0}\overline{\mathcal{L}}^{2}} \frac{\varepsilon_{\lambda} a_{0}}{\pi} \nonumber \\
 & =  &  \frac{f_p}{\Phi_{0}\varepsilon_{\lambda}} \left( \frac{U_{p}n_{i}}{\varepsilon_{0}}\right)\left(\frac{\Phi_{0}}{B}\right)^{1/2}. \hspace{4mm}(B \gg B^{*})
\label{current}
\end{eqnarray}
The main parameter, $\overline{\mathcal{L}}= \left( \pi U_{p}n_{i}/\varepsilon_{\lambda} \varepsilon_{0} \right)^{-1/2}$, is the average distance between effective defects pinning a single vortex in the low--field limit.\cite{vdBeek2002} The crossover field $B^{*}$ is that above which the intervortex repulsion limits the number of effective pins per vortex, and $f_{p}$ is the maximum pinning force exerted by a single strong pin. $U_p{\mathrm /[J]}$ is the pinning energy of a single strong pin, $n_i$ is the pin density, $a_{0} = (\Phi_{0}/B)^{1/2}$ is the intervortex distance, and $\varepsilon_{\lambda} = \lambda_{ab}/\lambda_{c}$ is the penetration depth anisotropy.

\begin{figure}[t]
\includegraphics[width=0.6\textwidth]{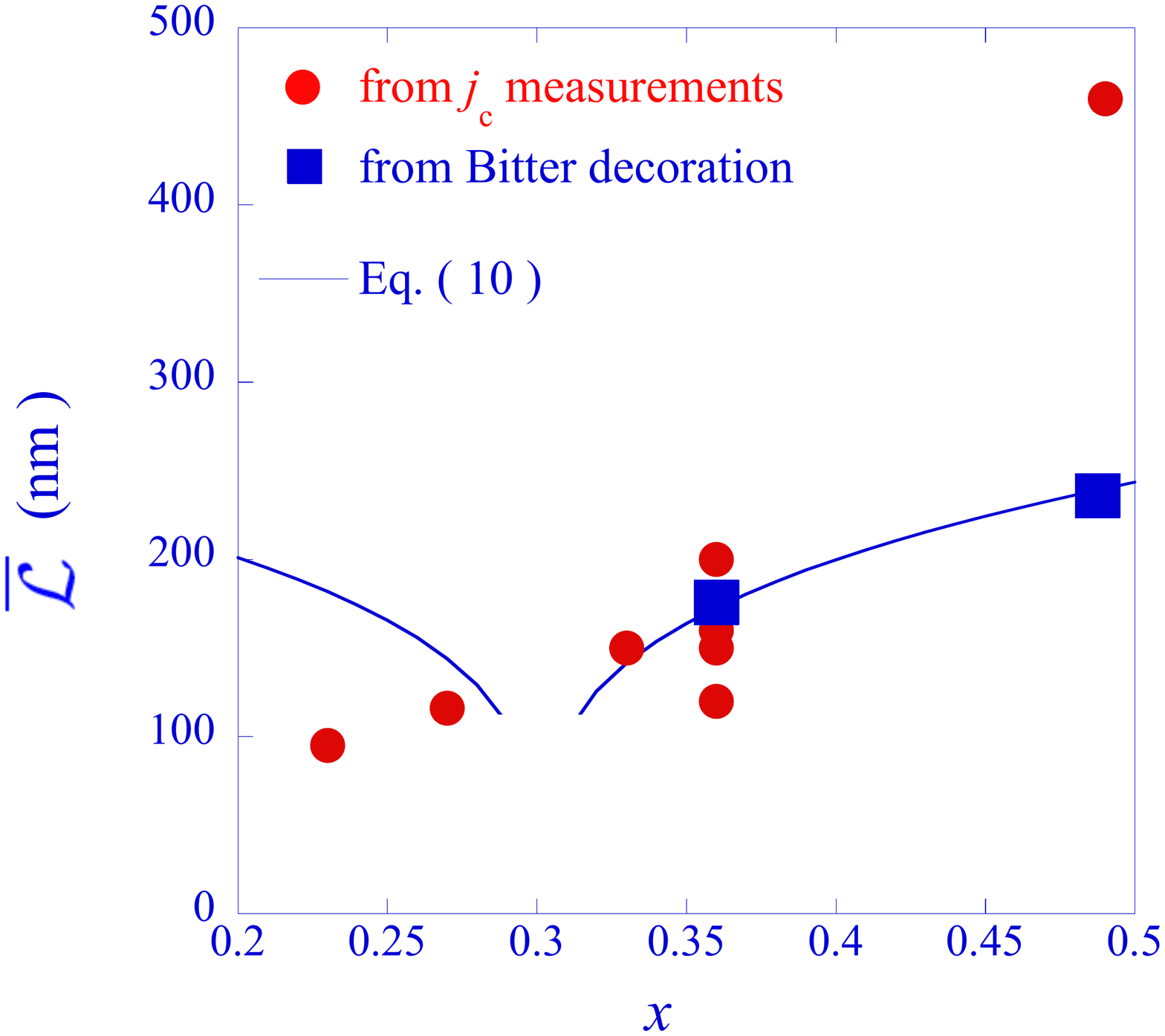}
\caption{(Color online) Average distance $\overline{\mathcal L}$ between effective pins versus P content. $\overline{\mathcal L}$ is obtained from $j_{s}(B)$--data at 5~K (red bullets \protect\color{red}$\bullet$\protect\color{black}), and from Bitter decoration at $T_{f} = 0.87 T_{c}$ (blue squares \color{blue}\protect\rule{2mm}{2mm}\color{black}). The drawn lines show the comparison with Eq.~(\protect\ref{eq:length-density-variations}), for variations of the dopant atom density $\Delta x = 0.3$~\%, on a characteristic length scale $\delta z = 100$~nm. }
\label{pinning}
\end{figure}

Equations~\eqref{current1} and \eqref{current}  show that the pinning force of a single strong pin  $f_{p} = ( \Phi_{0}^{3/2}\varepsilon_{\lambda} / \pi) \left\{ j^{2}(0)/[\partial j(B)/\partial B^{-1/2}]\right \}$ can be obtained from the experimentally measured low temperature, low-field  current density $j_{c}(0)$, and the slope $\partial j_{c}(B)/\partial B^{-1/2}$ at intermediate fields.  Reserving our attention to the crystals used in the Bitter decoration experiments presented below,  we obtain, for an estimated $\varepsilon_{\lambda} = 0.15$,\cite{Prozorov2009}  $f_{p} \approx 8 \times 10^{-13}$~N 
for both crystals ($x= 0.36 \#$2) and ($x= 0.49 \#$1).
This value is twice larger than that measured in Ba(Fe$_{1-x}$Co$_{x}$)$_{2}$As$_{2}$.\cite{demirdis} An evaluation at the highest measurement temperature of $0.8T_{c}$ yields $f_{p} = 2\times 10^{-14}$~N; however, this value is likely to be overestimated due to creep. Similarly, one can extract the length $\overline{\mathcal L} = f_{p}/\Phi_{0}j_{c}(0)$. Figure~\ref{pinning} shows that  $\overline{\mathcal{L}}$ is of the order of several dozen to hundreds of nm, in accordance with the strong pinning hypothesis. Moreover, the distance between effective pins clearly increases as function of P-content $x$.

 \begin{figure}[t]
\includegraphics[width=0.5\textwidth]{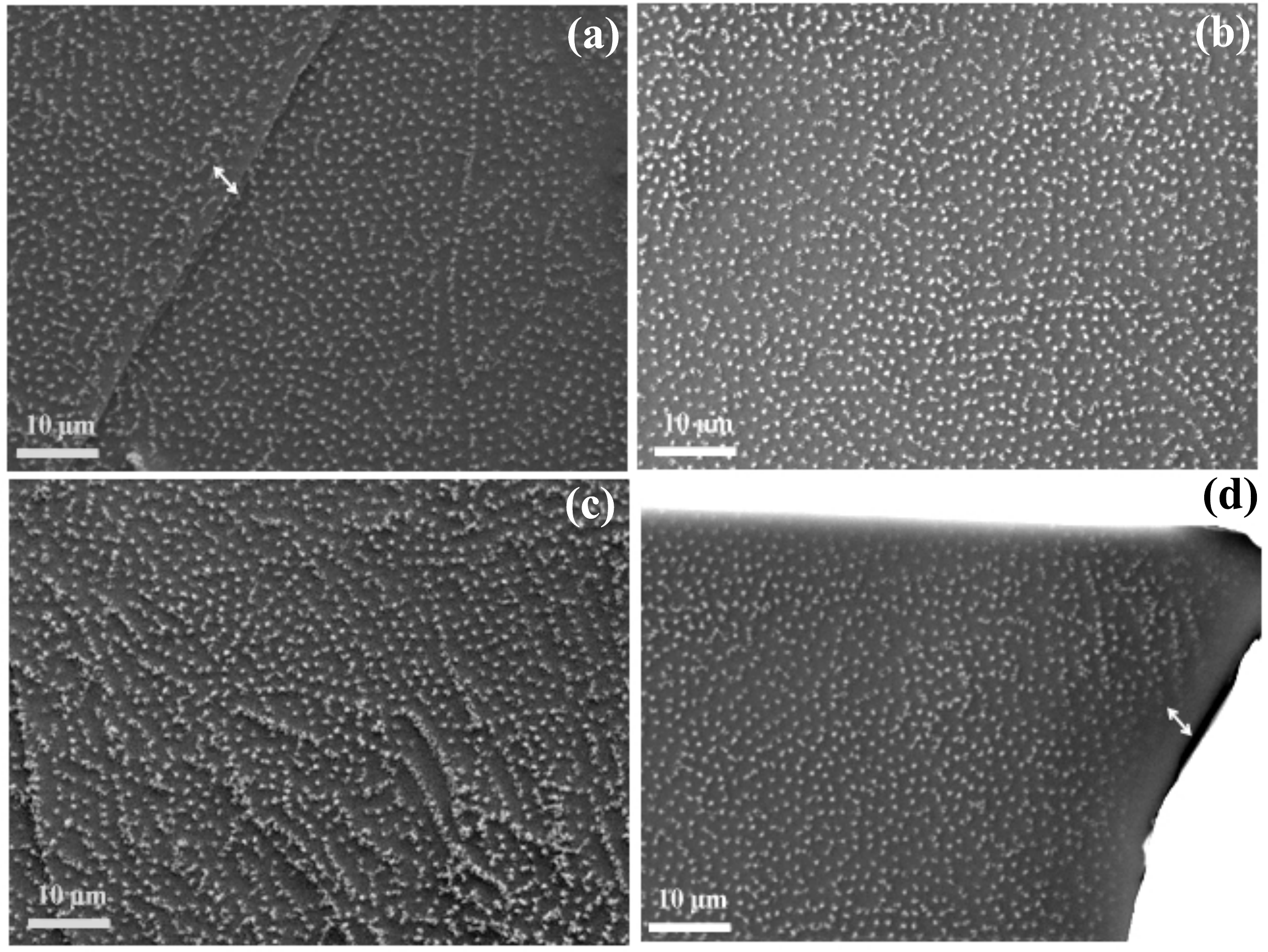}
\caption{Bitter decoration images of BaFe$_{2}$(As$_{1-x}$P$_{x}$)$_{2}$ single crystals for an applied field $\mu_{0}H_{a}$= 20~G (a,c) $x=0.36 \, \#2$, and (b,d) $x = 0.49\, \#1$. The white arrows  indicate the vortex-free Meissner belt observed (a)  near a surface step and (d) at the edge of the crystal $x=0.49\, \# 1$. The regions where images (b) and (d) were obtained with respect to the entire crystal are depicted in Fig.~\ref{DMO}. }
\label{deco}
\end{figure}

\subsection{Vortex imaging by Bitter decoration}
\label{section:Bitter} 

The Bitter decoration technique \cite{fasano,demirdis} was used to image  the vortex ensemble in BaFe$_{2}$(As$_{1-x}$P$_{x}$)$_{2}$ single crystals with three different substitution levels, $x=0.33$, $x=0.36$, and $x=0.49$.  The experiments were realized under  field cooled (FC) conditions, with a field $\mu_{0}H_{a} =20$~G (2 mT) applied parallel to the $c$-axis of the crystals. The decoration experiment for the crystal with $x=0.33$ was not successful, presumably due to the large value of the penetration depth.\cite{Hashimoto2012}  

 The vortex configurations shown in Fig.~\ref{deco}, for crystals $x=0.36\, \# 2$, and $x=0.49\,  \# 1$, are representative of what is observed over the entire specimens. From the decoration images, we obtain, for both crystals,  the average value of the magnetic induction as $B_{int}=n_{v} \Phi_{0}\approx 19$~G (with $n_{v}$  the vortex density).  This is 1~G smaller than the applied field during the experiment. Notwithstanding Ref.~\onlinecite{Prozorov2010}, there is therefore evidence for Meissner exclusion of the magnetic flux. Moreover, the Meissner current manifests itself as a vortex-free ``Meissner belt'' along the edges of  decorated crystal  $x=0.49 \, \# 1$, as well as near  surface steps that appear during preliminary cleavage of the samples, as indicated in Figure~\ref{deco}. Long vortex chains reminiscent of those observed in Ref.~\onlinecite{haihuwen} are also observed in the decoration images, for both investigated P-contents, however the chains are more pronounced in crystal $x=0.36 \, \#2$.

\begin{figure}[t]\vspace{-4mm}
\includegraphics[width=0.48\textwidth]{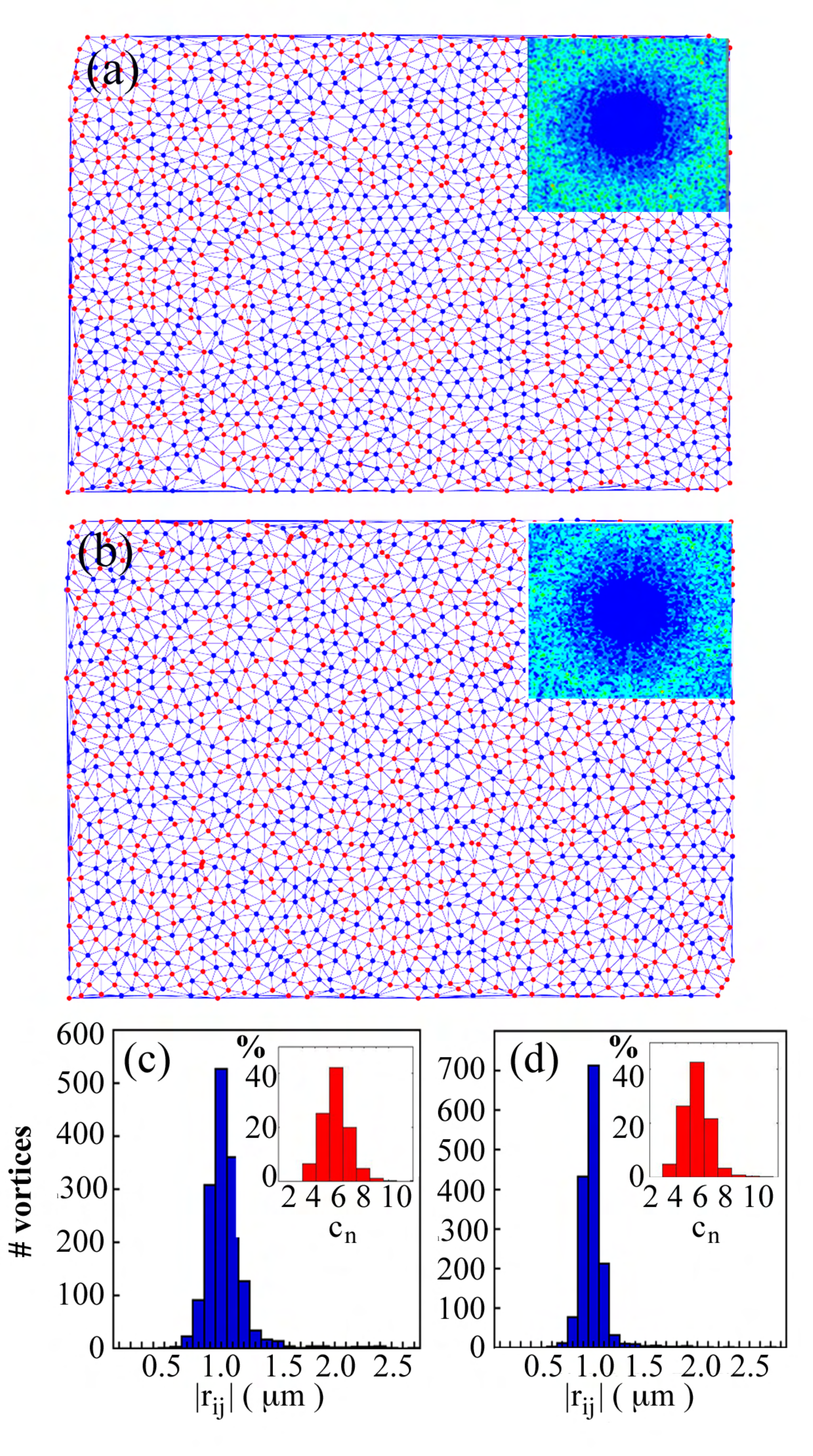}
\vspace{-10mm}
\caption{(Color online)  Delaunay triangulation of vortex ensembles observed in BaFe$_{2}$(As$_{1-x}$P$_{x}$)$_{2}$ single crystals (a) $x=0.36 \, \# 2 $ and (b) $x=0.49 \, \# 1$, and presented in Figs.~\ref{deco}\,(a,c) respectively.  Blue dots represent vortices with sixfold coordination, while red dots represent  differently coordinated vortices. The insets show the Fourier transform of the vortex positions. (c,d) Nearest neighbor distance distributions for the respective triangulations. The insets present the respective coordination number distributions. }
\label{TR}
\end{figure}
 
Figures\,~\ref{TR}\, (a) and (b) present the respective Delaunay triangulations of the vortex ensembles of Fig.~\ref{deco} (a) and (b). Here, blue dots represent  vortices with sixfold coordination, while red dots represent vortices with different coordination number.  Even if there are some regions with several adjacent sixfold coordinated vortices, the insets to Fig.\,~\ref{TR}\,(c) and (d) reveal that the latter represent less than half of the total  (46$\%$ for crystal  $x=0.49 \, \#1$, and 43$\%$ for crystal $x=0.36 \, \#2$). In fact, for the magnetic field of 20 G under study, the percentage of sixfold coordinated vortices is the same as in Ba(Fe$_{0925}$Co$_{0.075}$)$_{2}$As$_{2}$ (43\%) \cite{Mathieu} and in Ba$_{0.6}$K$_{0.4}$Fe$_{2}$As$_{2}$ (44\%).\cite{haihuwen} Moreover, the coordination number histograms have the same width for the three materials, revealing similar disorder of the vortex ensemble. The absence of vortex lattice order is further brought out by the Fourier transforms of the vortex positions shown in the insert to Fig.~\ref{TR} (a) and (b). Nevertheless, the spatial distribution of vortices in the two BaFe$_{2}$(As$_{1-x}$P$_{x}$)$_{2}$ crystals presents smaller density fluctuations than previously observed in the Co-doped material.\cite{demirdis} Panels (c,d) of Fig.~\ref{TR} show the distributions of nearest-neighbor intervortex distances. These have a mean value  $\lvert r_{ij}\rvert =1$~$\mu$m, while the lattice parameter for a triangular perfect  lattice of the same density is $a_{\triangle}=1.075 \sqrt{\Phi_0/B}=1.12 $~$\mu$m. This shift is due to the existence of more densely packed vortices,  notably in the chain-like structures. 

 \begin{figure}[t]
\includegraphics[width=0.49\textwidth]{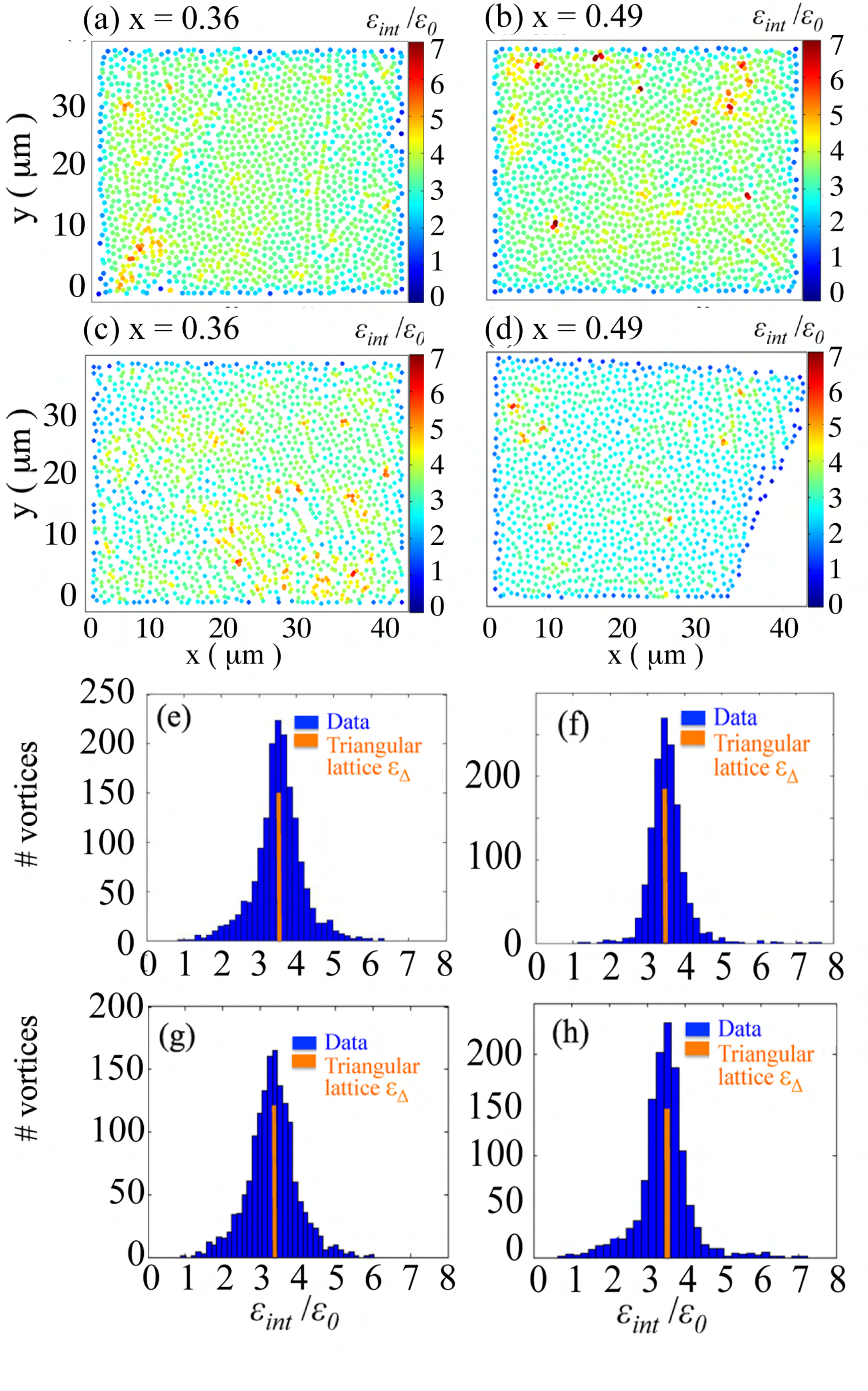}
\caption{(Color online) (a-d) Color-coded maps of the normalized individual vortex interaction energies calculated from the images of Fig.\,~\ref{deco}(a-d) using Eq.~(\protect\ref{Eint}), and represented in the same configuration.  Lower panels:  normalized interaction energy distributions for BaFe$_{2}$(As$_{1-x}$P$_{x}$)$_{2}$ crystals  ($x=0.36 \, \# 2$) (e,g), and ($ x=0.49 \, \# 1$) (f,h). Vortices located within a distance $10 \lambda_{ab}$ from the map edge are excluded from the histograms. The interaction energy per vortex of the triangular lattice ($\delta$-function) is represented by the central orange line in each histogram.}
\label{Emap}
\end{figure}

 \begin{figure}[t]
 \vspace{-3mm}
\includegraphics[width=0.45\textwidth]{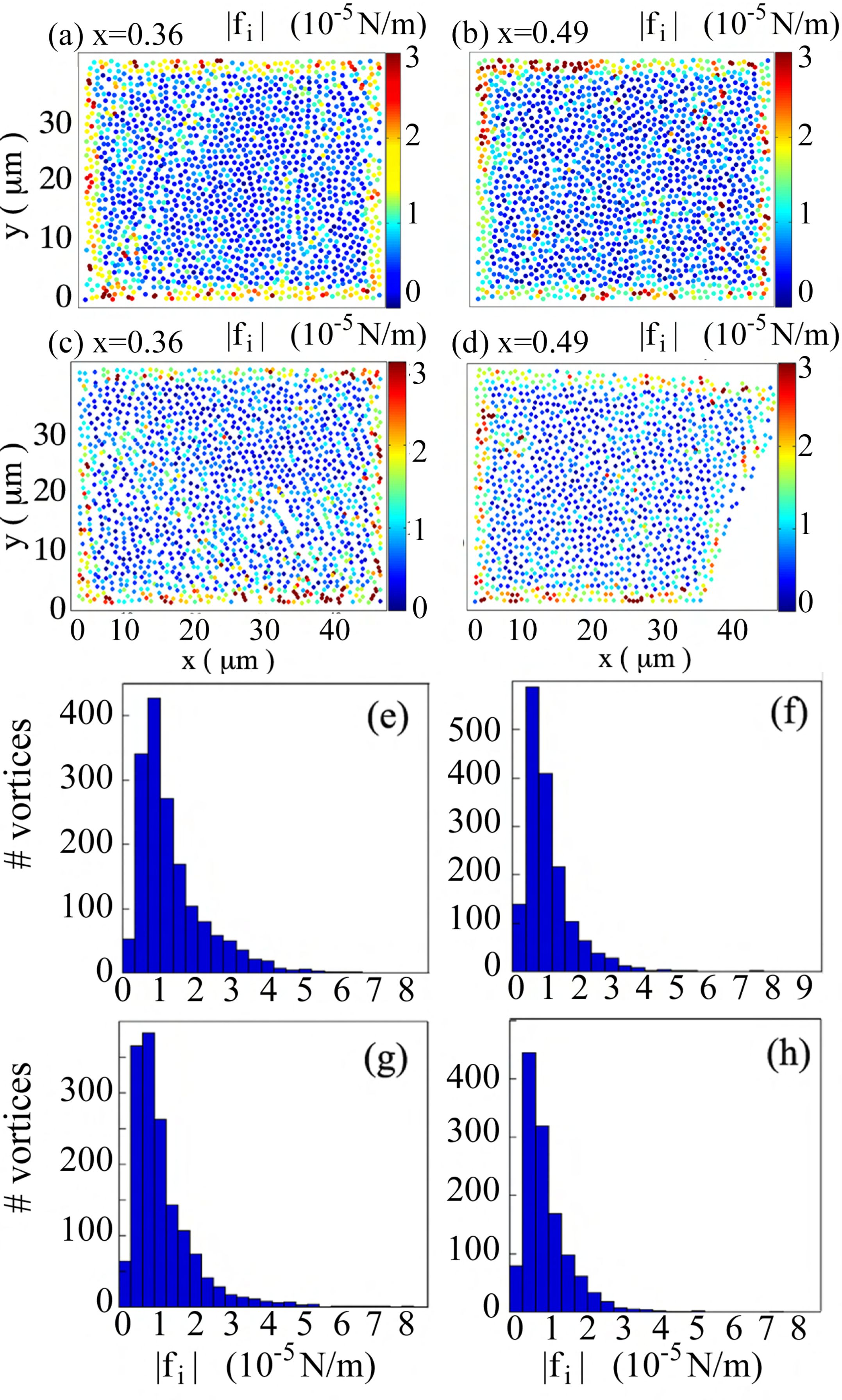}
\caption{(Color online) Normalized color-coded maps of the modulus of the pinning force (per unit length), calculated from the images of Fig.~\protect\ref{deco} using Eq.~(\protect\ref{fp}). Lower panels:  pinning force distributions for BaFe$_{2}$(As$_{1-x}$P$_{x}$)$_{2}$ single crystals ($x=0.36\, \# 2$),  (e,g), and ($x=0.49 \, \# 1$), (f,h). Again, vortices located within a distance $10 \lambda_{ab}$ from the map edge are excluded from the histograms. }
\label{Fmap}
\end{figure}

\section{Discussion}
\label{pinningforce}

In what follows, we adopt the procedure of Ref.~\onlinecite{demirdis} to determine the vortex interaction energy. For this, one needs to know the value of $\lambda_{ab}$ at the temperature $T_f$ at which the vortex ensemble is  frozen.  As in Ref.~\onlinecite{demirdis}, we use the information that can be obtained from vortex lines situated near surface steps. Such steps may act as barriers, but, due to the circulation of the Meissner current, they also prevent vortex lines from being situated right at their edge. Inserting the height of the surface step in Fig.~\ref{deco}\, (a), $h=1.3$~$\mu$m, and the width of the vortex-free region close to the step, $u=1.2$~$\mu$m, in Eq.~(1) of Ref.~\onlinecite{demirdis}, the value of the penetration depth at the freezing temperature is graphically estimated as $\lambda_{ab}(T_{f})\approx 700$~nm. Using the temperature dependence of $\lambda_{ab}(T)$ from Ref.~\onlinecite{Hashimoto}, one obtains the freezing temperature of the vortex ensemble as $T_{f} \approx 0.87 T_{c}$. Thus, even though vortices are frozen at a relatively high reduced temperature,  its value is lower than $T_{f}/T_{c} =0.95$ found in Ba(Fe$_{1-x}$Co$_{x}$)$_{2}$As$_{2}$.

Using  $\lambda_{ab}(T_{f})$ and the vortex positions extracted from Fig.\,~\ref{deco}, the interaction energies of the individual vortices can be calculated as 
\begin{equation}
\mathcal E_{int}^{i} = \sum_{j} 2\varepsilon_{0} K_{0}\left(\frac{|r_{ij}|}{\lambda_{ab}}\right).
\label{Eint}
\end{equation}
Here $K_{0}(x)$ is the lowest-order modified Bessel function, and $\lvert r_{ij}\rvert$ is the distance from vortex $i$ to  vortex $j$. For each vortex $i$, only neighbors $j$ contained within a circle of radius of $10 \lambda_{ab}(T)$ are taken into account. This radius was chosen after verification that vortices situated at a larger distance do not significantly contribute to ${\mathcal E}_{int}^{i}$.

Figures~\ref{Emap}(a-d) present the vortex interaction energies as color-coded maps, with the energy scale normalized by $\varepsilon_0(T_{f})$, as extracted from the images of Fig.~\ref{deco} (a) to (d) respectively.  The maps show a globally  homogeneous distribution; however,  a number of denser regions exist.  Histograms of the interaction energies for the maps (a) to (d) are presented in Fig.~\ref{Emap} (e) to (h). Note that the presence of the chain-like features with a denser vortex arrangement broadens the histograms for BaFe$_{2}$(As$_{1-x}$P$_{x}$)$_{2}$ single crystal ($x=0.36\,\# 2$). Still, the energy distributions are considerably narrower than those found in Ba(Fe$_{1-x}$Co$_{x}$)$_{2}$As$_{2}$.\cite{demirdis}   Furthermore, all distributions are centered about the average  $\overline{\mathcal E}_{int} \approx 3.5 \varepsilon_{0}$, which corresponds to the interaction energy value ($\delta$-peak) of the perfectly traingular Abrikosov lattice for this particular vortex density.  Therefore, in contrast to  Ba(Fe$_{1-x}$Co$_{x}$)$_{2}$As$_{2}$,\cite{demirdis}  no pinning-induced shift of the average value of the energy distribution histogram with respect to the $\delta$-peak value is observed.

\begin{figure}[t]
 \vspace{-3mm}
\includegraphics[width=0.45\textwidth]{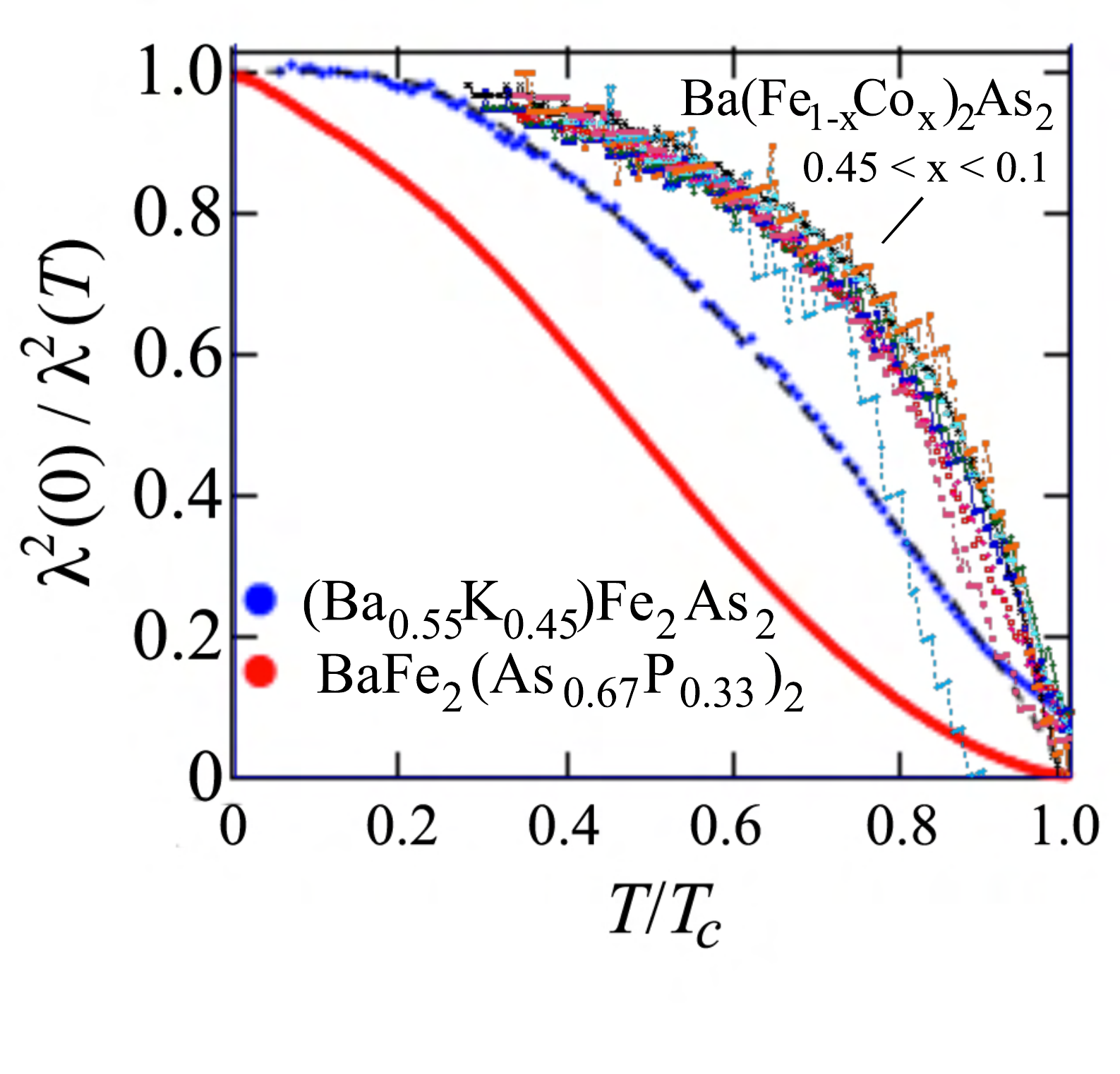}
 \vspace{-9mm}
\caption{Temperature dependence of the superfluid density, $n_{s}(T)/n_{s}(0) \propto \lambda_{ab}^{2}(0)/\lambda_{ab}^{2}(T)$ for various doping levels $x$ of  Ba(Fe$_{1-x}$Co$_{x}$)$_{2}$As$_{2}$ (upper curves, small data points), of Ba$_{0.55}$K$_{0.45}$Fe$_{2}$As$_{2}$ (\color{blue}$\bullet$\color{black}) and BaFe$_{2}$(As$_{0.67}$P$_{0.33}$)$_{2}$(\color{red}$\bullet$\color{black}) (both from Ref.~\protect\onlinecite{Hashimoto}).
 }
\label{fig:superfluiddensity}
\end{figure}

The large shift found in Ref.~\onlinecite{demirdis} was interpreted in terms of a large average pinning energy in the vicinity of $T_c$, which can only be the result of $T_c$ heterogeneity. The absence of such a shift in isovalently substituted BaFe$_{2}$(As$_{1-x}$P$_{x}$)$_{2}$ suggests that spatial inhomogeneity of $T_{c}$ is irrelevant  for vortex pinning in this material. A probable reason for this is a smoother temperature dependence of the pinning potential. The pinning energy due to spatial variations of the vortex line energy, on a scale $\delta z$ and with variance $\Delta \varepsilon_{0}$, writes $U_{p} \sim \Delta \varepsilon_{0}(T) \delta z\sim   [ \Delta n_{s}(0)/n_{s}(0)] [\partial \varepsilon_{0}(T)/\partial n_{s}(0)] n_{s}(T)+ [\varepsilon_{0}/n_{s}(0)] [\partial n_{s}(T) / \partial T_{c}] \Delta T_{c} $. At temperatures close to $T_{c}$, the second contribution, due to spatial fluctuations of $T_{c}$, dominates the pinning energy.\cite{demirdis}  However, whereas in materials such as Ba(Fe$_{1-x}$Co$_{x}$)$_{2}$As$_{2}$ -- in which  $n_{s}$ is nearly linear in $1-T/T_{c}$ as one approaches $T_{c}$ (see Fig.~\ref{fig:superfluiddensity})-- this contribution is large, in  BaFe$_{2}$(As$_{1-x}$P$_{x}$)$_{2}$ with a smoother temperature dependence of $n_{s}$ this contribution vanishes.  In fact, the $T_c$ heterogeneity within the decorated areas of the P-substituted material, such as observed by the DMO technique, does not result in  qualitatively different vortex arrangements in different parts of the crystal [see Figure~\ref{DMO}\,(b)].

Figure~\ref{Fmap} shows maps of the modulus of the  pinning force for each individual vortex, calculated from 
\begin{equation}
\mathbf{f}_{i} = \sum_{j} \frac{2\varepsilon_0}{\lambda_{ab}} \frac{\mathbf{r}_{ij}}{|\mathbf{r}_{ij}|} K_{1}\left(\frac{|\mathbf{r}_{ij}|}{\lambda_{ab}}\right)
\label{fp}
\end{equation}
following a similar procedure as used for the determination of $\mathcal{E}_{int}^{i}$. Here $K_{1}(x)$ is the first order modified Bessel function. Since the rendered vortex configurations in Fig.~\ref{deco} are in a stationary state at the freezing temperature $T_f$, the calculated intervortex repulsive force must be balanced by the pinning force.  The maps of Fig.~\ref{Fmap}(a) to (d) therefore represent the minimum local pinning force for each vortex, $\mathbf{ min}(|{\mathbf f}_{i}|)$. The distributions of $\mathbf{ min}(|{\mathbf f}_{i}|)$ shown in Fig.~\ref{Fmap}(e) to (h) allow one to estimate the average pinning force per vortex and per unit length. We obtain  ${|\overline{\mathbf{f}}_{i}|} \sim 3.5 \times 10^{-6}$~Nm$^{-1}$ for  crystal $x=0.49\,\#1$  and ${|\overline{\mathbf{f}}_{i}|}\sim 4.5 \times 10^{-6}$~Nm$^{-1}$ for  crystal $x=0.36\, \# 2$. These (high temperature) values are comparable to those found in Ba(Fe$_{1-x}$Co$_{x}$)$_{2}$As$_{2}$.\cite{demirdis}

The maps of the local pinning force moduli and the interaction energy shown in Figs.~\ref{Emap} and \ref{Fmap} respectively are correlated. However, the respective probability distributions are clearly broader for the lower substitution level $x = 0.36$. As in Ref.~\onlinecite{demirdis}, the ratio of the  elementary pinning force per pin $f_{p}$, extracted from the $j(B)$ curves in section~\ref{currentdensity}, and the value of ${|\overline{\mathbf{f}}_{i}|}$ obtained from Bitter decoration, allows one to evaluate an upper bound on  $\bar {\mathcal{L}}$  in an independent manner.
Using the low-temperature value $f_{p} \sim 8\times 10^{-13}$ N yields  $\bar{\mathcal{L}} = 180$ nm for  crystal ($ x=0.36\, \#2$) and  $\bar{\mathcal{L}} = 230$~nm for  crystal ($x=0.49 \, \#1$). Figure~\ref{pinning} shows that these numbers are consistent with  those directly extracted from the sustainable current density measurements. 

Several  reasons can be invoked to explain the observed enhancement of $\overline{\mathcal L}$ with increasing P-content $x$. First is the increase of the superfluid density $n_{s}(x)$ as function of $x$.\cite{Hashimoto2012}  Other possibilities are a decrease of the density $n_{i}$ of pinning centers for larger $x$, and the decrease of the penetration depth anisotropy $\varepsilon_{\lambda}^{-1}$ for higher substitution levels. We first investigate the effect of the increase of the superfluid density with $x$. Assuming that nm-scale  fluctuations of the dopant atom density (with variance $\Delta x$) are spatially isotropic, their effect on the pinning force 
\begin{equation}
\left \langle f_{p} \right \rangle  \sim   \langle \int_{\delta z} \nabla \varepsilon_{0}(r) dz \rangle \sim \Delta \varepsilon_{0}
\end{equation}
and the pinning energy $U_{p} \sim \Delta \varepsilon_{0} \delta z$ can be estimated by exploiting the dependence $\lambda_{ab}(x)$,\cite{Hashimoto2012} 
\begin{equation}
\langle f_{p}  \rangle \sim \Delta \varepsilon_{0} \sim \frac{\partial \varepsilon_{0}}{\partial \lambda_{ab}} \frac{\partial \lambda_{ab}}{\partial x}  \Delta x.
\end{equation}
The low--field critical current density (\ref{current1}) becomes 
\begin{eqnarray}
j_{c} & \sim &  \frac{\pi^{1/2} n_{i}}{\Phi_{0}\varepsilon_{\lambda}} \frac{\Delta \varepsilon_{0}^{3/2}}{\varepsilon_{0}^{1/2}} \nonumber  \\
& = & \frac{n_{i}}{\Phi_{0}\varepsilon_{\lambda}} \left( \frac{\pi}{\varepsilon_{0}(x)}\right)^{1/2}\left|\frac{\partial \varepsilon_{0}}{\partial \lambda_{ab}}  \frac{\partial \lambda_{ab}}{\partial x} \Delta x  \right|^{3/2};
\label{eq:dopant-density-variations}
\end{eqnarray}
the length between effective pinning centers
\begin{eqnarray}
\overline{\mathcal L} & \sim & \frac{\varepsilon_{\lambda}}{\pi^{1/2}} \left( \frac{\varepsilon_{0}}{n_{i} \Delta \varepsilon_{0} \delta z }\right)^{1/2} \nonumber \\
& =& \frac{\varepsilon_{\lambda}}{2 \pi^{1/2}} \frac{1}{(n_{i} \delta z )^{1/2} } \left|\frac{1}{\varepsilon_{0}}\frac{\partial \varepsilon_{0}}{\partial \lambda_{ab}}  \frac{\partial \lambda_{ab}}{\partial x} \Delta x  \right|^{-1/2}.
\label{eq:length-density-variations}
\end{eqnarray}
Assuming that for density fluctuations of the dopant atoms the pin density $n_{i} \sim (\delta z)^{-3}$ scales as the inverse cube of the defect size, these expressions can be directly compared to the dependence of the critical current density on P-content (see Fig.~\ref{Tc}c), as well as that of $\overline{\mathcal L}$ (see Fig.~\ref{pinning}). For $\delta z = 100$~nm and $\Delta x = 0.3$~\%, the qualitative trend  with $x$ of both quantities can be reproduced. Thus,  the dependence of the penetration depth on P-content accounts, at least in the overdoped regime, for an enhanced probability of encountering larger critical current densities around optimal substitution. The similar dispersion of critical current density and pinning length data for different $x$ around the model estimations (\ref{eq:dopant-density-variations}) and (\ref{eq:length-density-variations}) in  Figs.~\ref{Tc}c and~\ref{pinning} would then be due to similar disorder of the P--distribution in the different growth batches, eliminating the need to invoke a decreasing disorder with increasing $x$. Still, the evolution $\overline{\mathcal L}(x)$ is reminiscent of that of the normal-state mean free-path, extracted by Shishido {\em et. al.} from de Haas-van Alphen oscillations of the Landau magnetization.\cite{Shishido} They reported  that the mean free-path for the $\beta$ orbits in BaFe$_{2}$(As$_{1-x}$P$_{x}$)$_{2}$ single crystals increases from $l \sim 170$~ \AA \, to 800~\AA \, when P content  varies from $x= 0.41$ to 1. This would imply the presence of structural defects that act as strong pinning centers that are unrelated to spatial composition fluctuations. Last but not least, the increase of $\varepsilon_{\lambda}(x)$, as inferred from the doping-dependent evolution of the Fermi surface,\cite{Shishido} and corresponding to a smaller Fe-pnictogen distance for large $x$, would lead to stiffer vortex lines and to less pinning for higher P-content.

Finally, we address the clear presence of  chain-like structures in the Bitter decoration images. The following hypotheses can be suggested to explain the appearance of these chains. The first is the possibility of native heterogeneity of the crystals under study, introduced during growth. The presence of line-like defects, or of linear agglomerates of point--like defects giving stronger local pinning, will lead to the appearance of vortex alignments or chainlike structures, much as this was found in {\em e.g.} Bi$_{2}$Sr$_{2}$CaCu$_{2}$O$_{8+\delta}$.\cite{Oral97,Fasano99,Fasano2000,Soibel2001,MingLi2004} A second possibility is that, the vortex images having been obtained after field--cooling, and since the BaFe$_{2}$(As$_{1-x}$P$_{x}$)$_{2}$ material under consideration show Meissner expulsion, a certain fraction of vortices must exit the material before the vortex ensemble is frozen at $T_{f}$. The chains possibly correspond to flow channels for these exiting vortices, the flux in intermediate areas remaining pinned.\cite{Olson98} Finally, the multi-band character of BaFe$_{2}$(As$_{1-x}$P$_{x}$)$_{2}$ may be responsible for the occurrence of  vortex chains, with the possible existence of an attractive part in the inter-vortex potential.\cite{Brandt-Type1-5} We leave these questions open for further work.

 \section{Summary and Conclusion}
 
We have presented an overview of vortex pinning in single crystals of the isovalently substituted iron-based superconductor BaFe$_{2}$(As$_{1-x}$P$_{x}$)$_{2}$, in which we have attempted to correlate the sustainable screening current density as a function of temperature, field, and doping $x$, with the structural properties of the vortex ensemble. The critical current density in BaFe$_{2}$(As$_{1-x}$P$_{x}$)$_{2}$ is, overall, very well described by the strong pinning scenario of Ref.~\onlinecite{vdBeek2002}, which allows one to extract elementary pinning forces (of the order of $10^{-13}$ N) and the distance between effective pins. The latter is of the order of 100~nm, and increases as a function of doping level $x$. These values are consistent with those independently obtained by means of the magnetic decoration technique. Contrary to Ref.~\onlinecite{Fang}, we find no contribution of weak collective pinning to the sustainable current density, suggesting that P-atoms are not responsible for quasi-particle scattering.\cite{Kees2} The sustainable current data are affected by flux creep, which prohibits one from drawing definite conclusions concerning the temperature dependence of pinning.

Bitter decoration reveals slightly more ordered vortex ensembles (the number of sixfold coordinated vortices is slightly higher) than those observed in charge--doped Ba(Fe$_{1-x}$Co$_{x}$)$_{2}$As$_{2}$. Also, the interaction energy and pinning force distributions in  BaFe$_{2}$(As$_{1-x}$P$_{x}$)$_{2}$ are much narrower than those in Ba(Fe$_{1-x}$Co$_{x}$)$_{2}$As$_{2}$, and are not shifted with respect to the interaction energy of a perfectly triangular vortex lattice with the same density.  These observations exclude a role of spatial variations of the critical temperature $T_{c}$ in determining the frozen vortex state obtained upon field--cooling. The absence of the weak-collective pinning contribution to the critical current density in  BaFe$_{2}$(As$_{1-x}$P$_{x}$)$_{2}$ means that the strong pinning contribution is what generates the disordered  vortex configurations.

The main features of strong vortex pinning in BaFe$_{2}$(As$_{1-x}$P$_{x}$)$_{2}$, such as the energy-- and force histograms, the density of effective pins, and the evolution of the critical current density with P-content, can be understood using a model for heterogeneity of the superfluid density on the scale of several dozen to several hundred nanometers, due to an inhomogeneous distribution of the dopant atoms. A small spatial variance  $\Delta x$ ( of the order of 0.3~\% )  suffices to explain the magnitude of $j_{c}$. This model explains why pinning is of extrinsic origin, and why the disorder can be readily annealed.\cite{furukawa}  Even if the $x$--dependence of the critical current density can be well described without any assumption of less disorder in crystals with higher P-content, the reminiscence of the evolution with P-content of the mean distance between effective pinning sites and the normal-state mean-free path\cite{Shishido} suggests that a second type of pinning centers may be at play. This could be areas of enhanced local strain, more prominent at optimal doping, such as these arise from the very different Fe-As and Fe-P distances.\cite{rotter}

\section*{Acknowledgements} 
This work was made possible thanks to the support of the ECOS-Sud-MINCyT France-Argentina bilateral program, Grant No. A09E03. Work done in Bariloche was partially funded by PICT 2007-00890 and PICT 2008-294.

\end{document}